\documentclass[12pt,a4paper]{article}

\usepackage[british]{babel}

\usepackage[a4paper,top=2cm,bottom=2cm,left=2.5cm,right=2.5cm,marginparwidth=1.75cm]{geometry}

\usepackage{amsmath}
\usepackage{graphicx}
\usepackage[colorlinks=true, allcolors=blue]{hyperref}
\usepackage{orcidlink}
\usepackage[title]{appendix}
\usepackage{textcomp}
\usepackage[utf8]{inputenc}
\usepackage{newunicodechar}
\newunicodechar{₂}{$_2$}
\usepackage{mathrsfs}
\usepackage{amsfonts}
\usepackage{booktabs} 
\usepackage{caption}  
\usepackage{threeparttable} 
\usepackage{algorithm}
\usepackage{algorithmicx}
\usepackage{algpseudocode}
\usepackage{listings}
\usepackage{enumitem}
\usepackage{chngcntr}
\usepackage{lipsum}
\usepackage{subcaption}
\usepackage{authblk}
\usepackage[T1]{fontenc}    
\usepackage{csquotes}       
\usepackage{diagbox}
\usepackage{siunitx}
\usepackage{array}

\usepackage{cite}

\usepackage{setspace}
\usepackage{titlesec}
\titleformat{\section} 
  {\normalfont\Large\bfseries}{\thesection.}{1em}{}
  
\usepackage{float}   


\title{\textbf{Neutron-Induced Enhancement of Ion Transport Through Lithium-Ion Battery Materials}}

\author[1,2,*,**]{Ha M. Nguyen}
\author[1,3,**]{Carson D. Ziemke}
\author[5]{David Stalla}
\author[5]{Tymofii Pieshkov}
\author[6]{Thi-Hien Truong}
\author[5]{Min Su}
\author[2]{Bikash Saha}
\author[2]{Narendirakumar Narayanan}
\author[1,3]{Carlos Wexler}
\author[2]{John Gahl}
\author[1,4,*]{Yangchuan Xing}

\affil[1]{\small Materials Sciences and Engineering Institute, University of Missouri, Columbia, MO 65201, USA}
\affil[2]{University of Missouri Research Reactor (MURR), University of Missouri, Columbia, MO 65203, USA}
\affil[3]{Department of Physics and Astronomy, University of Missouri, Columbia, MO 65201, USA}
\affil[4]{Department of Chemical and Biomedical Engineering, University of Missouri, Columbia, MO 65201, USA}
\affil[5]{Electron Microscopy Core, University of Missouri, Columbia, Missouri, USA}
\affil[6]{Advanced Materials Characterization Core Facility, Missouri University of Science and Technology, Rolla, Missouri, USA}

\affil[*]{Corresponding authors: \texttt{hn4gq@missouri.edu, xingy@missouri.edu, heitmannt@missouri.edu}}
\affil[**]{Authors with equal contributions}

\begin{document}
\maketitle

\begin{abstract}
Polycrystalline solid-state ionic conductors (SSICs) are essential energy materials for all-solid-state Li-ion batteries. To date, achieving a room-temperature ionic conductivity of solid electrolytes comparable to that  of their liquid counterparts remains a critical challenge. Here, we experimentally demonstrate that thermal neutron irradiation can offer an innovative strategy in that neutron-induced modification in an SSIC model (LiBO$_{2}$ as an effective cathode coating) can facilitate ion transport through the material, enhancing its ionic conductivity. The central concept is that high-flux ($\sim 10^{9}\text{ neutrons}\cdot \text{cm}^{-2}\cdot \text{s}^{-1}$) thermal neutrons ($\sim \text{25 meV}$) selectively transmute strong neutron absorbers [which are $^{10}$B (3840 barns) and $^{6}$Li (940 barns) isotopes and present in their natural  abundances of $\sim 19.9\%$ and $\sim 7.5\%$, respectively, in polycrystalline grains of LiBO$_2$] to generate lattice vacancies without compromising their crystallographic long-range order. In addition, by-product gamma photons emitted from $^{10}$B transmutation free electrons to stop atomic displacement and simultaneously neutralize the space charge built up by positively-charged oxygen vacancies at grain boundaries. As a result, the ionic conductivity is increased by nearly 20\% for the grains and more than 80\% for the grain boundaries. This study validates theoretical predictions and highlights a vital strategy for boosting ion transport in ionic solids. Overall, this novel approach establishes a new revenue for broader applications and greater enhancements of advanced functional materials in their related solid-state ionic devices, including all-solid-state lithium-ion batteries. 

\end{abstract}
\section{Introduction}
 The term solid state ionics was first introduced by Takehiko Takahashi in 1967 to describe the study of solid state ionic conductors (SSICs) and their technological applications \cite{OsamuYamamoto2017,KlausFunke2013,HansRickert1978,Kazuya_Terabe_2022}. These include inorganic polycrystalline materials that are capable of conducting electricity by transporting mobile charge carriers, involving ions, through grains and grain boundaries \cite{HansRickert1978,EricaTruong2025}. If the charge carriers are pure ions, SSICs are classified as fully ionic conductors. Solid electrolytes (SEs) are examples of these conductors \cite{AtulKumarMishra2021}. In contrast, if SSICs possess both ionic and electronic carriers (\textit{e.g.}, electrons and electronic holes), they are mixed ionic-electronic conductors (MIECs) \cite{Riess2003,Sunarso2008,AbdulkadirKızılaslan2025} such as battery electrodes \cite{Ghosh2022,Hou2023, Xu2012,Hayner2012,Koech2024,BaoYuanyuan2025} and their coatings \cite{Guan2020,UmairNisar2021,Kaur2022,Tan2020,Maske2024,Chen2010, HaMNguyen2015,HaMNguyen2025a,HaMNguyen2025b,Guo2023,DWANG2015,Du2019,Gao2021,Li_2025,Zhang2019,Ramkumar2020}. Regardless of whether a SSIC functions as a SE or a MIEC, fast ion transport is central to its effective performance in solid-state ionic devices \cite{Kazuya_Terabe_2022,Li_and_Xiao_2022}, including sensors \cite{Garzon_2004, Kleitz_1986}, detectors \cite{Thomas_Defferriere_2022,Thomas_Defferriere_2024,Thomas_Defferriere_2025}, actuators \cite{Terabe_2022}, ionic-migration-based resistive switching devices (resistive random access memories and memristors) \cite{YangyinChen2020,SuhasKumar2022}, electrochemical capacitors (supercapacitors) \cite{XiaojunZeng_2019}, fuel cells \cite{JavedRehman2024}, and batteries \cite{MuhammadFaizan2025,book_advanced batteries}. In this work, we focus on engineering ion transport in functional SSICs for their effective application in Li-ion batteries (LIBs), with the goal of enhancing their ionic conductivity. 

The ionic conductivity $\sigma$ and the diffusivity $D$ in crystalline solids are related via the Nernst--Einstein relation \cite{Was_book_2007,HelmutMehrer_book_2007,Riviere_bookchapter_1995,Hayes&Stoneham_book_2004} , such that
\begin{equation}
\sigma = \frac{n q^{2} D}{k_{\text{B}} T},
\label{eq1}
\end{equation}
where $n$ is the concentration of mobile charge carriers, $q$ is the ionic charge, $k_{\text{B}}$ is the Boltzmann constant, and $T$ is the absolute temperature. Accordingly, $\sigma$ is exponentially dependent on both the entropy of thermal activation $\Delta S$ and the enthalpy of thermal activation $\Delta H$ (also referred to as the activation energy $E_{\text{a}}$ or migration energy barrier $E_{\text{m}}$) of diffusional processes \cite{Almond_1987,Yong_jun_Zhang_2012}, following
\begin{equation}
\sigma \propto \exp\!\left(\frac{\Delta S}{k_{\text{B}}}\right)\exp\!\left(-\frac{\Delta H}{k_{\text{B}} T}\right).
\label{eq2}
\end{equation}

In principle, Eq. (\ref{eq2}) shows that $\sigma$ can be exponentially enhanced by increasing $\Delta S$ while keeping $\Delta H$ constant, decreasing $\Delta H$ while keeping $\Delta S$ constant, or simultaneously increasing $\Delta S$ and decreasing $\Delta H$. Nevertheless, despite the extensive body of work devoted to chemical and structural defect engineering in SSICs, most established approaches rely on equilibrium or near-equilibrium synthesis strategies such as solid-state reaction methods in which defect populations are constrained by thermodynamic formation energies \cite{YuanRen_2026,Yanlong_Wu_2024,Mi2024,Xuyong_Feng_2022,HuangxuLi_2025,ShengyangDong_2025,JincanLi_2025,Zeng2022,ChuanWang_2024,Deck_Hu_2023,Zhang_Nazar_2022,Dhattarwal_2024,Chen_2022,Xiong_Wang_2026,Zhipeng_Wang_2025,Shuo_Wang_2023,Brandon_Wood_2021}. For example, Li-ion transport can be enhanced through several strategies, including the paddle-wheel mechanism, in which rotational motion of local anions \cite{Zhang_Nazar_2022} or electron pairs \cite{Dhattarwal_2024} facilitates diffusional jumps of Li$^{+}$ cations and thereby increases their diffusivity; introducing multiple cations and/or anions (e.g., a Li$^{+}$-stuffing strategy to enhance Li$^{+}$–Li$^{+}$ repulsion) \cite{Chen_2022}; incorporating additional components into the host SSIC in a cocktail-like fashion \cite{Xiong_Wang_2026,Zhipeng_Wang_2025}; and introducing chemical, structural, or dynamical frustration—i.e., a physical condition in which numerous competing states exist with comparable energies \cite{Shuo_Wang_2023,Brandon_Wood_2021}. In general, these approaches aim to increase the entropy of host SSICs, thereby promoting high-entropy SSICs with enhanced ion transport \cite{HuangxuLi_2025,ShengyangDong_2025,JincanLi_2025,Zeng2022,ChuanWang_2024}.

However, Deck and Hu stated in their recent review \cite{Deck_Hu_2023} that entropy-driven disorder improves Li$^{+}$ ion transport mainly by weakening local Li$^{+}$–anion interactions, but high macroscopic ionic conductivity emerges only when this local disorder is integrated into thermodynamically stable long-range structures that enable percolating diffusion pathways. As a result, introducing a sufficiently high concentration of transport-enhancing lattice defects—particularly in chemically and structurally robust oxides—remains intrinsically challenging \cite{Roman_Schlem_2021}. This limitation is especially pronounced for defect species with large formation energies, such as boron vacancies in LiBO$_2$ (the formation energy of boron vacancies is approximately 10 eV \cite{HaMNguyen2025a}), which have been theoretically predicted to substantially lower Li-ion migration barriers [$E_{\text{m}}$ = 0.27 eV for the monoclinic LiBO$_2$ polymorph (\textit{i.e.}, the $\alpha$ phase) and $E_{\text{m}}$ = 0.13 eV for the tetragonal LiBO$_{2}$ polymorph (\textit{i.e.}, the $\gamma$ phase)] \cite{HaMNguyen2025a}, yet are inaccessible through conventional chemical substitution or thermal processing techniques.

An alternative way to overcome these thermodynamic constraints is to employ non-equilibrium methods capable of selectively generating lattice disorder while preserving long-range crystallographic order \cite{HaMNguyen2025a,HaMNguyen2025b,Roman_Schlem_2021}. Particle irradiation represents one such approach, as it enables the dose-controlled introduction of lattice defects, defect clusters, and electronic excitations, etc. \cite{Kinchin_Pease_1955, Olsen, Kuhnke, MuhammadMominurRahman2020, tan2016,LingLi2018,Islam2012, Ogorofnikov2010, Haseman2021,Qiu2015}. In particular, thermal neutron irradiation offers a unique and largely unexplored opportunity for defect engineering in SSICs, owing to its ability to induce selective nuclear transmutation reactions in specific isotopes naturally present or  isotopically enriched in a concentration-controlled manner within the crystal lattice \cite{Shlimak1999,Logan2020,Barber2020,Tanenbaum_Mills1961,Polyakov2010,Kuriyama2002,Lee2011,Satoh1990,Elena_Echeverria_2024}.

\begin{figure*}
 \centering
        \includegraphics[width=1\textwidth]{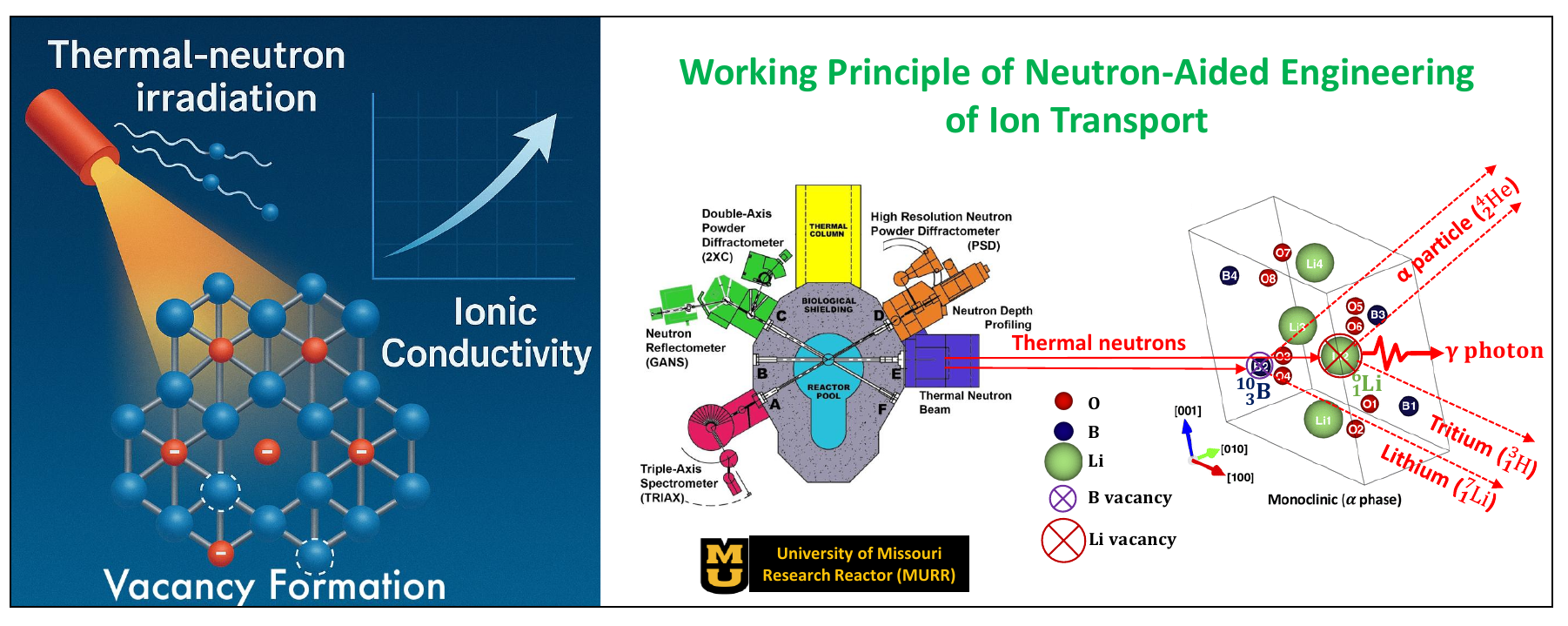}
 \caption{Working principle of thermal-neutron-irradiation approach to engineering of ion transport in LiBO$_2$.}
\label{figure1_working_principle}
\end{figure*}

Here, we demonstrate for the first time that thermal neutron irradiation can serve as an effective and selective strategy to enhance ion transport in LiBO$_2$, a model material for SSICs that has been shown to function effectively as a cathode coating for both liquid- and solid-electrolyte LIBs \cite{HaMNguyen2025a,HaMNguyen2025b,Guo2023,DWANG2015,Du2019,Gao2021,Li_2025,Zhang2019,Ramkumar2020} (see Refs. \cite{HaMNguyen2025a,HaMNguyen2025b} for a detailed review of recent studies on this material). Briefly, experimental reports have shown enhanced electrochemical performance and capacity retention in LiBO$_{2}$-coated LiNi$_{0.8}$Mn$_{0.1}$Co$_{0.1}$O$_{2}$ cathodes, namely, 90.1\% after 100 cycles at 1 C (3.0–4.6 V) \cite{DWANG2015}, 82.1\% after 100 cycles at 1 C and 4.5 V \cite{Du2019}, and 84.3\% after 150 cycles at 0.5 C \cite{Zhang2019}. Furthermore, doping boron into the cathode and coating it with LiBO$_{2}$ further enhances the performance, as shown by Gao \textit{et al.} \cite{Gao2021}, who demonstrated significant improvements in cycling stability and capacity retention under various conditions. Interestingly, Aykol \textit{et al.} \cite{Muratahan_Aykol_2016} has developed a screening framework based on high-throughput density functional theory that incorporates thermodynamic stability, electrochemical stability, and hydrofluoric-acid scavenging metrics to evaluate more than 130,000 oxygen-containing materials as potential cathode coatings, supporting the aforementioned experiments by predicting that LiBO$_{2}$ is among the best hydrofluoric-acid scavenging cathode coatings, namely, Sc$_{2}$O$_{3}$, Li$_{2}$CaGeO$_{4}$, LiBO$_{2}$, Li$_{3}$NbO$_{4}$, Mg$_{3}$(BO$_{3}$)$_{2}$, and Li$_{2}$MgSiO$_{4}$.

In LiBO$_2$, the presence of such strong neutron absorbers as $^{6}$Li (940 barns) and $^{10}$B (3840 barns) isotopes provides a natural platform for neutron-induced engineering of ion transport in crystalline solids \cite{Kinchin_Pease_1955,Elena_Echeverria_2024}. The working principle for this approach is provided in Figure \ref{figure1_working_principle}. The central concept is that high-flux thermal neutrons, harnessed at Beam Port E (rather than within the reactor core) of the University of Missouri Research Reactor (MURR), selectively induce nuclear transmutation of the $^{10}$B and $^{6}$Li isotopes, which uniformly occupy the crystal lattice sites of LiBO$_2$ with their natural isotopic abundances of $\sim 19.9\%$ and $\sim 7.5\%$, respectively. These transmutation reactions,
\begin{equation}
^{10}\text{B} + n \;  \rightarrow\; 
\begin{cases}
^{7}\text{Li} \,(1.015~\text{MeV}) + ^{4}\text{He} \,(1.777~\text{MeV}) & (6\%) \\
^{7}\text{Li} \,(0.840~\text{MeV}) + ^{4}\text{He} \,(1.470~\text{MeV}) + \gamma \,(0.480~\text{MeV}) & (94\%)
\end{cases}
\label{eq3}
\end{equation}

\noindent and 
{
\begin{equation}
^{6}\text{Li} + n \;\rightarrow\; ^{3}\text{H} \,(2.73~\text{MeV}) + ^{4}\text{He} \,(2.05~\text{MeV}),
\label{eq4}
\end{equation}
}

\noindent which shall be discussed in detail subsequently in Section~\ref{R&D}, generate lattice vacancies while preserving long-range crystallographic order. In addition, gamma photons produced as by-products of $^{10}$B transmutation \cite{Kinchin_Pease_1955,Olsen,Elena_Echeverria_2024,Walker_1960} liberate bonded electrons that mitigate atomic displacement through electron stopping mechanisms and simultaneously neutralize space-charge regions arising from positively charged oxygen vacancies at grain boundaries \cite{Thomas_Defferriere_2024}. Consequently, ionic conductivity increases by nearly 20\% within grains and by more than 80\% across grain boundaries.

Beyond LiBO$_2$, the approach presented here could establish a general framework for nuclear engineering of defects in SSICs, enabling access to defect configurations that are thermodynamically inaccessible via conventional synthesis methods. This strategy opens new avenues for simultaneously tuning enthalpic and entropic contributions to ion transport and provides a powerful complement to existing chemical and structural engineering methodologies for advanced solid-state ionic devices, including LIBs.

\section{Results and Discussions \label{R&D}}

\begin{figure*}
 \centering
        \includegraphics[width=0.7\textwidth]{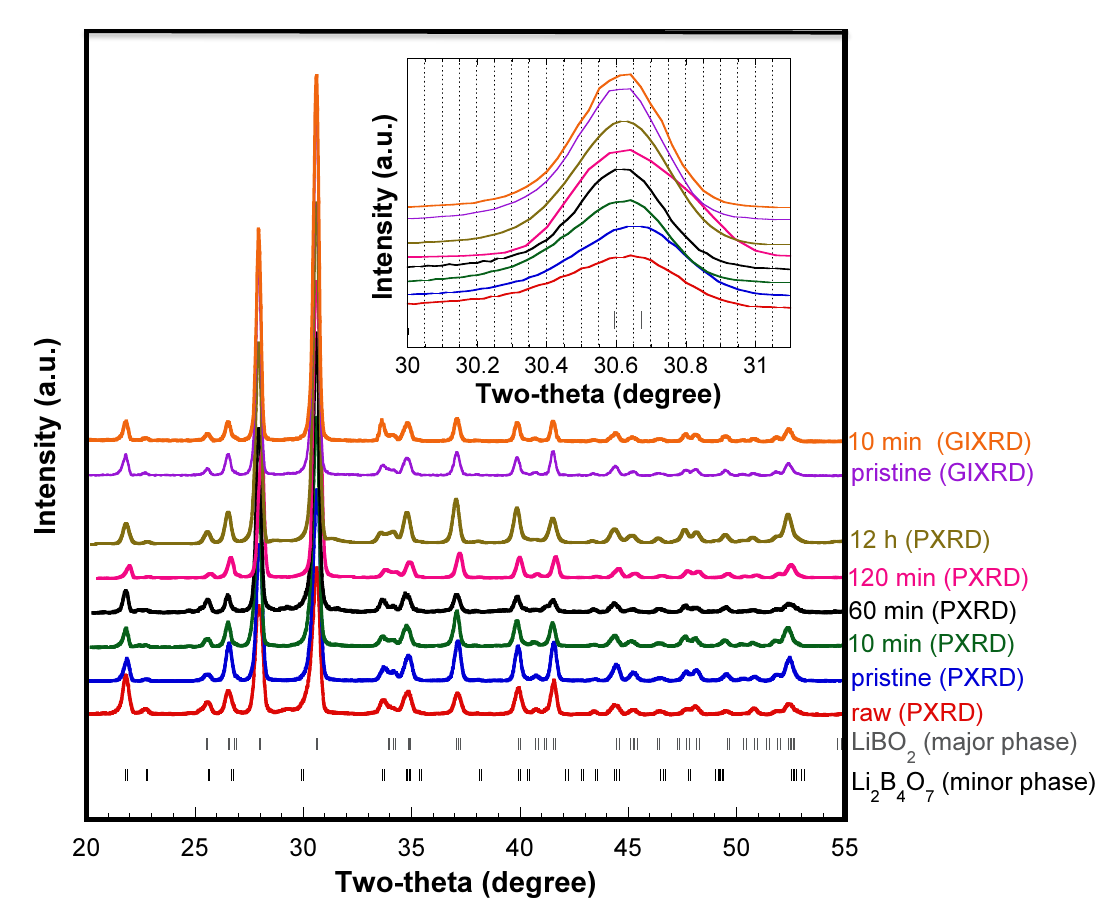}
 \caption{Powder X-ray diffraction (PXRD) and grazing incidence X-ray diffraction (GIXRD) patterns of raw, pristine, and irradiated samples of LiBO$_{2}$. The inset shows the intensive peak of LiBO$_{2}$ at around $2\theta \approx 30.6^{\text{o}}$}
\label{figure2_XRD}
\end{figure*}

Figure~\ref{figure2_XRD} shows the powder X-ray diffraction (PXRD) pattern (characterized on a Rigaku Ultima IV X-ray diffractometer) of the raw LiBO$_2$ powder. Rietveld refinement using the FullProf Suite indicates that the major phase (97\%) is monoclinic LiBO$_2$ (space group P2$_1$/c), with lattice parameters $a = 5.83$~\AA, $b = 4.34$~\AA, $c = 6.44$~\AA, $\alpha = \gamma = 90^\circ$, and $\beta = 114.96^\circ$. A minor secondary phase (3\%) corresponding to tetragonal Li$_2$B$_4$O$_7$ (space group I4$_1$cd) is also detected, with lattice parameters $a = b = 9.43$~\AA\ and $c = 10.28$~\AA (See also Table \ref{table1_rietveld}).

The PXRD pattern of powders obtained by grinding pristine pellets after sintering is also shown in Figure~\ref{figure2_XRD}. Rietveld refinement again reveals monoclinic LiBO$_2$ as the dominant phase, with lattice parameters $a = 5.86$~\AA, $b = 4.37$~\AA, $c = 6.48$~\AA, $\alpha = \gamma = 90^\circ$, and $\beta = 115.04^\circ$. A minor tetragonal Li$_2$B$_4$O$_7$ phase is also identified, with $a = b = 9.44$~\AA\ and $c = 10.45$~\AA (See also Table \ref{table1_rietveld}). At this stage, no additional crystalline phases are detected within the resolution of the Rietveld refinement. These results confirm that the sintering process preserves the monoclinic LiBO$_2$ phase of interest for subsequent characterization and neutron irradiation (see Section \ref{methods} for a detailed description of experimental methods and materials). 

\subsection{Structural and Microstructural Comparison: Pristine versus Low-Dose Irradiation \label{section3_1}}

\begin{figure*}
\centering
\includegraphics[width=\textwidth]{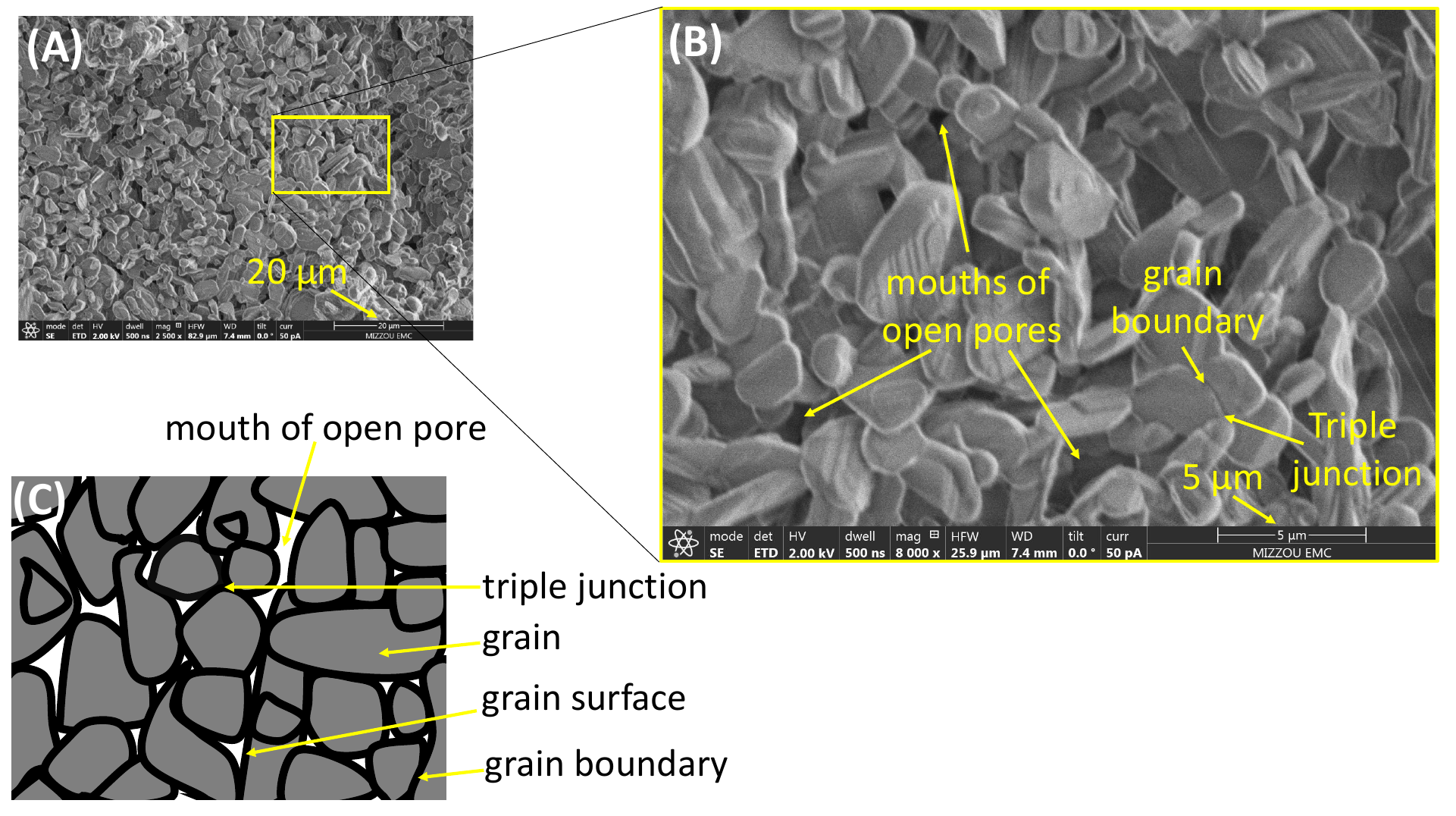}

\vspace{0.5em} 

\includegraphics[width=\textwidth]{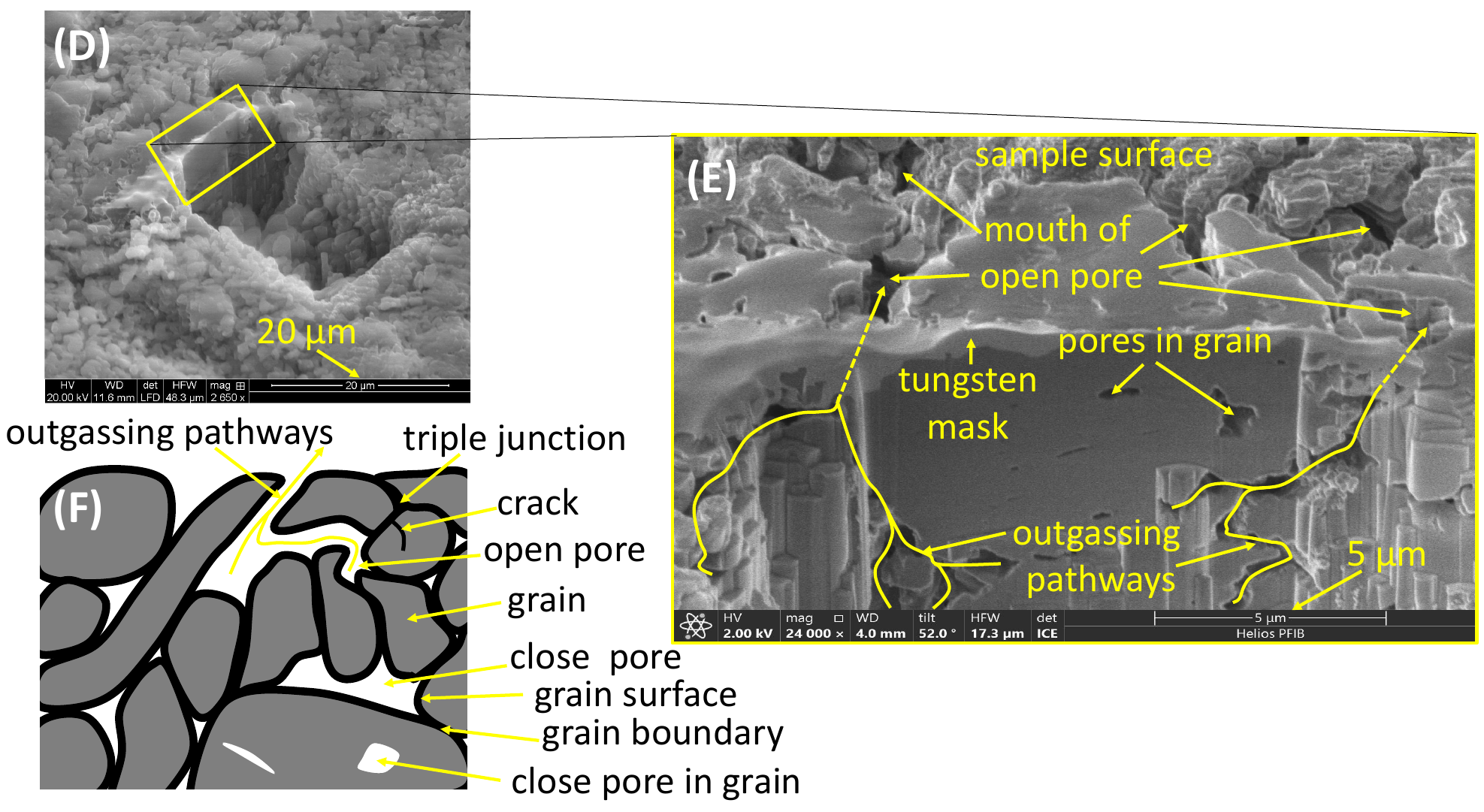}
\caption{Microstructure of pristine LiBO$_2$: SEM images (A, B, C) and FIB-SEM images (D, E, F).}
\label{figure3_SEM_pristine}
\end{figure*}

Figure \ref{figure3_SEM_pristine} presents the microstructural features of pristine polycrystalline LiBO$_{2}$ pellets revealed by scanning electron microscopy (SEM) and focused ion beam scanning electron microscopy (FIB-SEM) analyses. Low-magnification SEM imaging (panel \ref{figure3_SEM_pristine}A) shows a dense polycrystalline microstructure composed of irregularly-shaped gravel-like grains with characteristic sizes on the order of several micrometers. Higher-magnification SEM images (panel \ref{figure3_SEM_pristine}B) resolve well-defined grain boundaries and frequent triple junctions, together with the mouths of open pores located predominantly at grain-boundary intersections. These features are schematically illustrated in panel \ref{figure3_SEM_pristine}C to clarify the geometric relationships among grains, grain boundaries, triple junctions, and open pores.

Cross-sectional FIB-SEM images (panels \ref{figure3_SEM_pristine}D,E) further reveal the subsurface microstructure, highlighting interconnected pores and open pores extending from the surface into the interior of the pellet. Notably, continuous pathways associated with grain boundaries, triple junctions, and microcracks are observed, forming potential outgassing or transport channels through the polycrystalline network. The schematic in panel \ref{figure3_SEM_pristine}F summarizes these features, distinguishing between open and closed pores and grain-boundary regions. Importantly, no evidence of abnormal grain growth, large cracks, or fragmented domains within grains is observed in the pristine sample, indicating that the sintering process produces a mechanically intact and well-connected polycrystalline framework.

Overall, pristine LiBO$_{2}$ exhibits a dense polycrystalline microstructure comprising well-connected grains with clearly defined grain boundaries, triple junctions, and a limited amount of residual porosity. These microstructural features are expected to play a critical role by providing structurally well-defined transport pathways that can be selectively altered by subsequent neutron irradiation, as discussed below. Consequently, a detailed understanding of the structure and microstructure of the pristine LiBO$_{2}$ is essential for interpreting the irradiation-induced modifications and the associated transport phenomena examined in this work.

\begin{figure*}
\centering
\includegraphics[width=\textwidth]{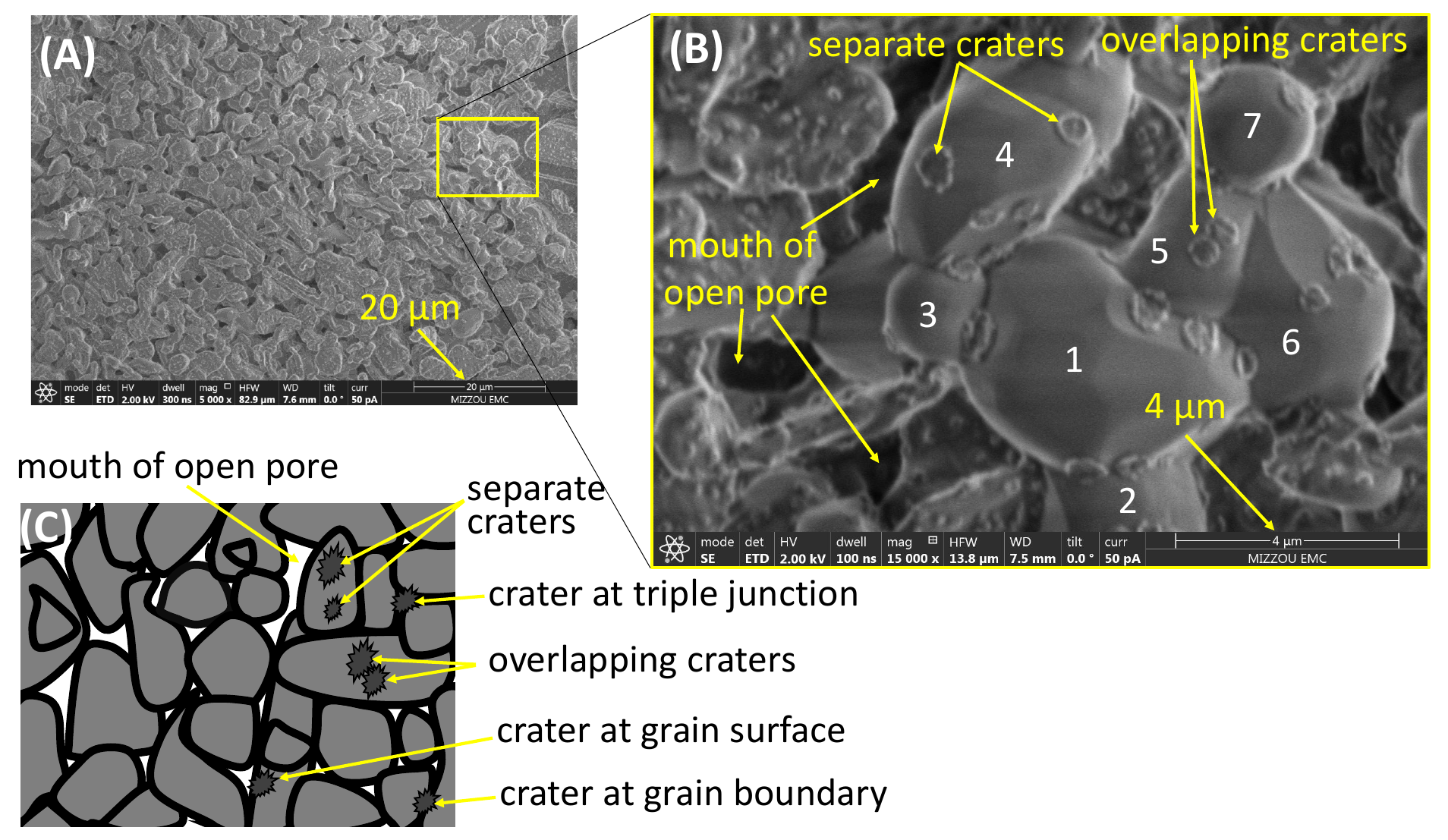}

\vspace{0.5em}

\includegraphics[width=\textwidth]{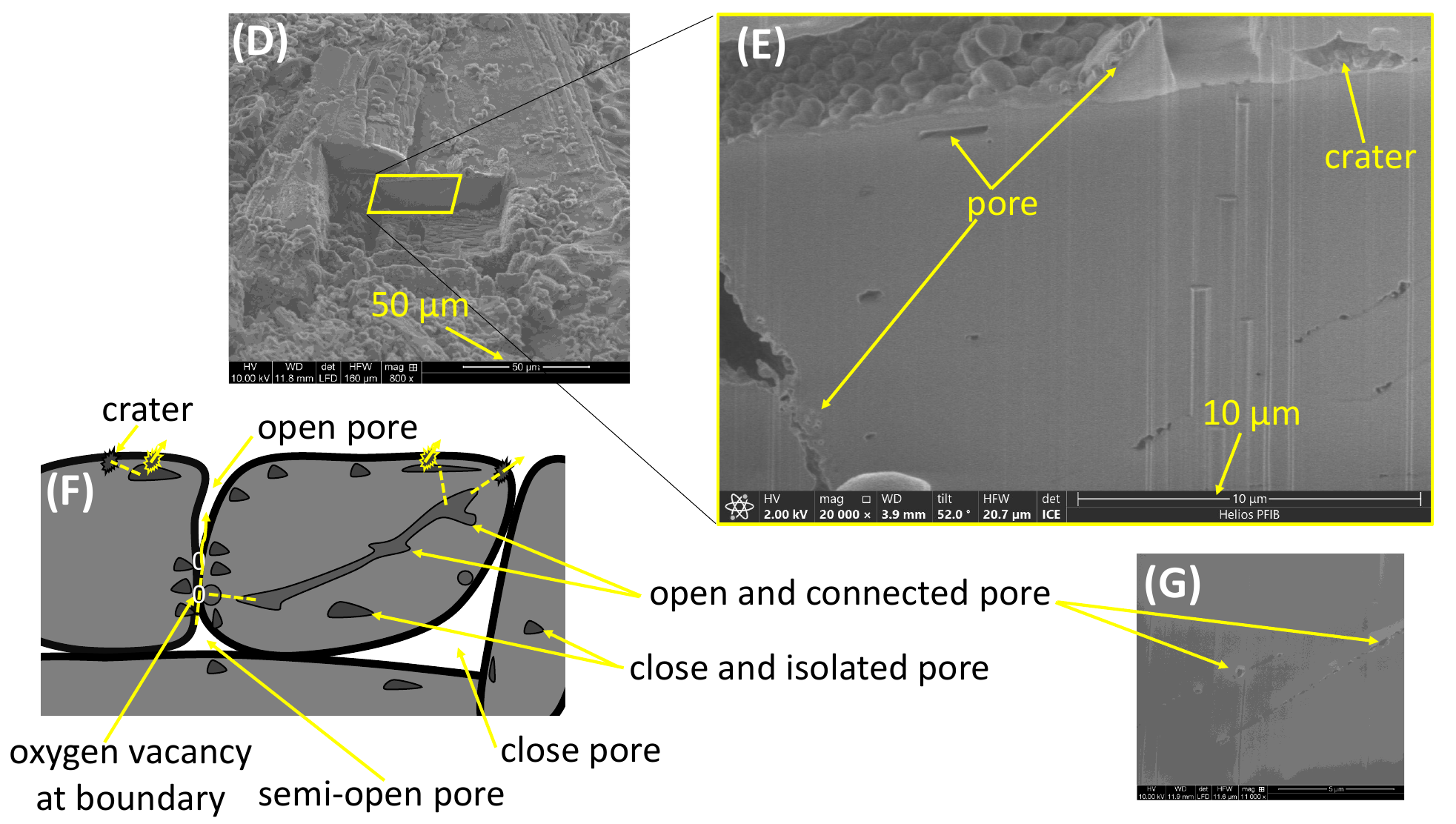}
\caption{Microstructure of 10 min irradiated LiBO$_2$: SEM images (A, B, C) and FIB-SEM images (D, E, F, G).}
\label{figure4_SEM_10min_irradiated}
\end{figure*}

Figure~\ref{figure4_SEM_10min_irradiated} presents the microstructural evolution of LiBO$_2$ pellets after 10~minutes of thermal neutron irradiation, as revealed by SEM and FIB--SEM analyses. Low-magnification SEM imaging (panel~\ref{figure4_SEM_10min_irradiated}A) shows that the overall polycrystalline framework remains intact, with no evidence of macroscopic cracking or grain fragmentation. At higher magnification (panel~\ref{figure4_SEM_10min_irradiated}B), however, numerous crater-like surface features are observed across grain surfaces, grain boundaries, and triple junctions.

The numbered grains (1--7) in panel~\ref{figure4_SEM_10min_irradiated}B reveal pronounced grain-to-grain variability in both crater density and morphology. Certain grains (e.g., grains 5 and 6) exhibit multiple overlapping craters, whereas adjacent grains (e.g., grains 1 and 4) show comparatively fewer or more isolated features. This spatial heterogeneity indicates that crater formation is governed by local microstructural factors, such as crystallographic orientation, grain-boundary character, and proximity to pre-existing pores. In several cases, craters on neighboring grains appear spatially correlated across grain boundaries, suggesting cooperative microstructural evolution at triple junctions and interconnected pore networks rather than independent, isolated events.

These crater-like features occur both as isolated depressions and as overlapping structures and are frequently associated with the mouths of pre-existing open pores, indicating that neutron irradiation preferentially amplifies existing surface and subsurface defects. The schematic in panel~\ref{figure4_SEM_10min_irradiated}C summarizes the spatial distribution of these irradiation-induced craters, emphasizing their preferential formation at microstructural heterogeneities such as grain boundaries, triple junctions, and open-pore interfaces. 

The formation of crater-like surface features can be attributed to highly localized energy deposition immediately after the thermal neutron capture by $^{6}$Li and $^{10}$B via nuclear transmutation reactions. These reactions produce energetic charged particles---tritons ($^{3}$H$^{+}$), alpha particles ($^{4}$He$^{2+}$), and lithium ions ($^{7}$Li$^{+}$)---with kinetic energies spanning the keV to MeV range. In this regime, energy loss is governed by coupled electronic and nuclear stopping processes, which can be described within the Bethe--Bloch formalism at high velocities and by nuclear stopping theory at lower velocities.

As these ions propagate through the solid, their stopping power increases with decreasing velocity, leading to a pronounced maximum in energy deposition near the end of their trajectories (Bragg peak). Specifically, tritons with energies of $\sim$2.7~MeV exhibit projected ranges on the order of 30--100~$\mu$m~\cite{RosaMariaMontereali_2023}, whereas alpha particles with energies of 1.5--2.0~MeV have shorter ranges of $\sim$5--15~$\mu$m due to their higher charge state~\cite{Pocaterra_2022}. In contrast, recoil lithium ions, with typical energies around 0.9~MeV, are dominated by nuclear stopping and therefore have much shorter ranges of $\sim$0.5--2~$\mu$m~\cite{Kondo_2025}. 

In this terminal regime, where nuclear stopping dominates, dense displacement cascades are initiated over characteristic length scales of tens to hundreds of nanometers. Primary knock-on atoms (PKAs) generated in these events transfer energy through successive collisions, producing large numbers of atomic displacements. The number of displaced atoms can be estimated using the Kinchin--Pease relation:
\begin{equation}
N \approx 0.8 \frac{E}{E_d},
\label{eq5}
\end{equation}
where $E$ is the PKA energy and $E_d$ is the displacement threshold energy (typically $\sim$20--50~eV for oxides). These cascades evolve on ultrafast timescales ($\sim 10^{-13}$--$10^{-11}$~s), during which transient thermal spikes and localized pressure buildup occur, creating a highly disordered or transiently molten region. For example, a recoil lithium ion ($^{7}$Li$^{+}$) produced in the transmutation of $^{10}$B with a kinetic energy of $\sim$0.8--1.0~MeV can generate on the order of $10^{3}$--$10^{4}$ displaced atoms according to Eqn.~(\ref{eq5}). 

Assuming a characteristic atomic spacing of $\sim$1.5~\AA, the corresponding atomic number density is on the order of $n \sim (1.5~\text{\AA})^{-3} \approx 3 \times 10^{22}~\text{cm}^{-3}$. The volume of a single cascade region can therefore be estimated as $V \sim N/n$, yielding $V \sim 3 \times 10^{-20}$ to $3 \times 10^{-19}~\text{cm}^3$. Approximating the cascade as a sphere, the corresponding radius is $r \sim (3V/4\pi)^{1/3}$, resulting in a characteristic primary cascade size of only $\sim$2--5~nm. 

\begin{figure*}
\centering
\includegraphics[width=\linewidth]{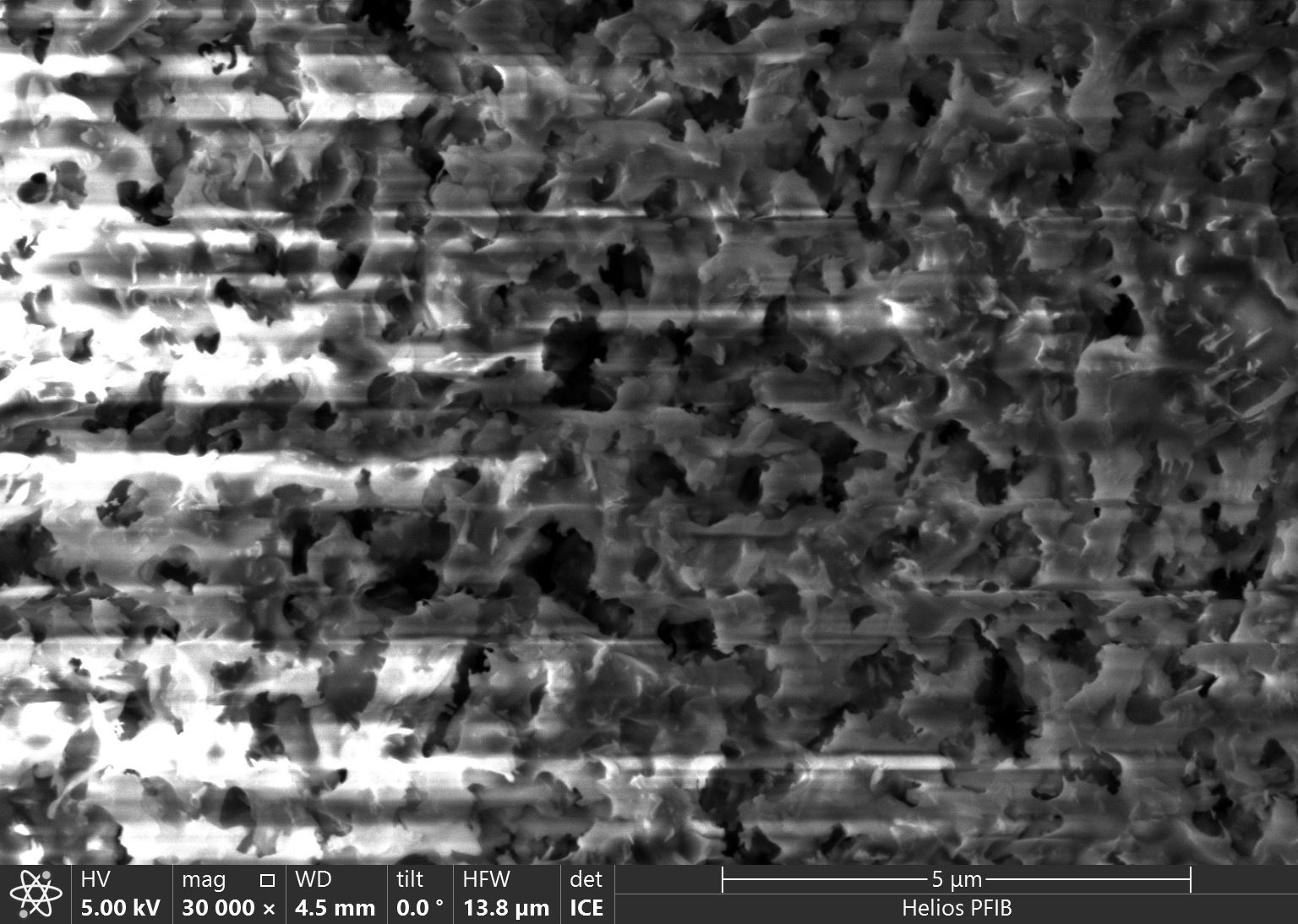}
\caption{High-resolution scanning electron microscopy image of microstructure inside a crater on the surface of 10 min irradiated LiBO$_2$. Craters formed inside a crater are clearly observed.}
\label{craters_in_craters}
\end{figure*}

Because an individual displacement cascade ($\sim$2--5~nm) is two-to-three orders of magnitude smaller than the micrometer-scale craters observed in SEM, these mesoscale features cannot represent single nuclear impact events. Given the incident neutron flux ($10^{9}\text{ neutrons}\cdot\text{s}^{-1}\cdot\text{cm}^{-2}$) over a 10-minute irradiation duration ($600\text{ s}$), the cumulative surface exposure yields an average arrival frequency of one incident neutron per square centimeter per nanosecond. Over 10 minutes, multiple spatial and temporal nuclear transmutation events occur in close proximity. The resulting subcascades and localized thermal spikes overlap, leading to a superposition effect: a "craters-within-a-crater" hierarchy wherein sub-micron damage zones coalesce into the observed micrometer surface depressions. This hierarchy can be seen vividly in Figure \ref{craters_in_craters}, which is a high-resolution SEM micrograph sampled from a large crater generated on the surface of a grain of the 10-min irradiated LiBO$_{2}$ pellet.

Cross-sectional FIB--SEM images (panels \ref{figure4_SEM_10min_irradiated}D,E) provide further insight into the subsurface response to irradiation. Beneath surface craters, extended pores, partially open pores, and interconnected pores are observed, indicating that neutron irradiation promotes the growth and interconnection of pre-existing pores rather than inducing uniform bulk damage. Continuous porous pathways extending from the surface into the interior are evident in some regions, while other pores remain isolated, as schematically illustrated in panel \ref{figure4_SEM_10min_irradiated}F. Grain interiors remain largely dense and structurally coherent, with no widespread cracking observed (panel \ref{figure4_SEM_10min_irradiated}G).

The PXRD and GIXRD results (Figure \ref{figure2_XRD} and Table \ref{table1_rietveld}), together with SEM and FIB--SEM observations, confirm that 10 minutes of thermal neutron irradiation preserves the overall long-range monoclinic matrix without bulk phase transformations or peak broadening. 

In the next subsection, the "craters-within-a-crater" surface morphology and isolating real neutron-induced defect tracks from environmental degradation or electron-beam artifacts are investigated with transmission electron microscopy (TEM) in order to further resolve their fine nanoscale structural hierarchy.


\subsection{Nanoscale Structure and Microstructure Resolution of Low-Dose Irradiated LiBO$_{2}$ using Transmission Electron Microscopy \label{section3_2}}

\begin{figure*}
\centering
\includegraphics[width=\linewidth]{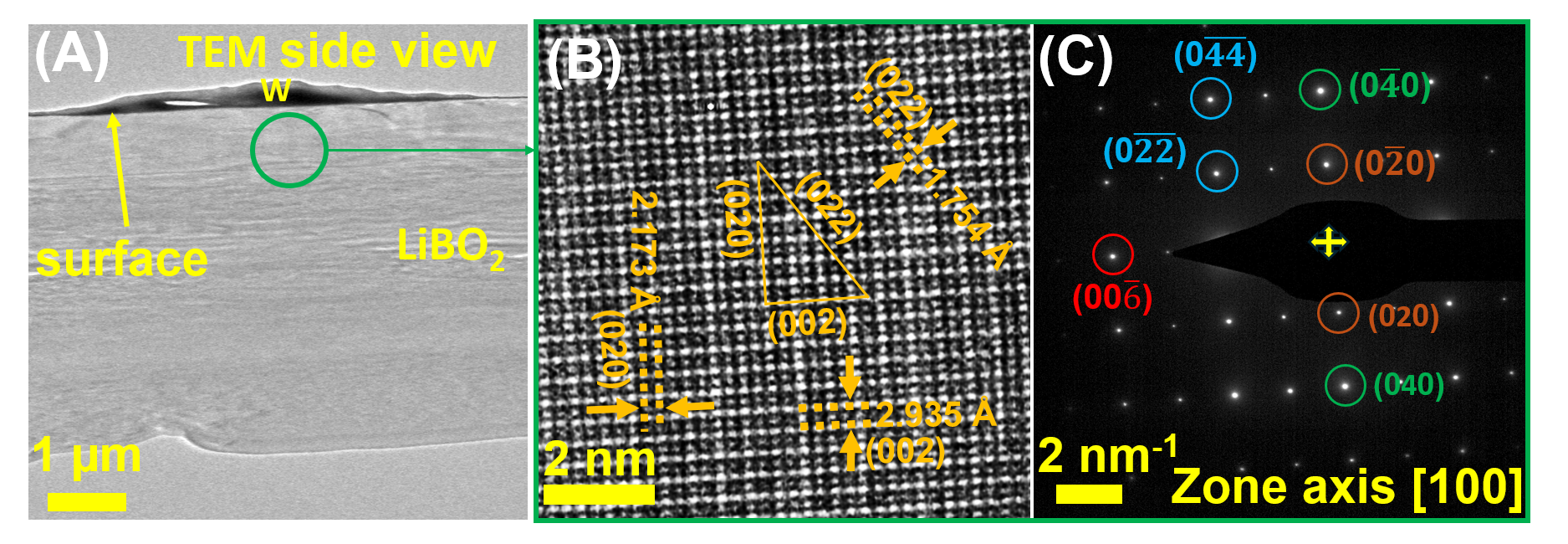}
\caption{Transmission electron microscopy analysis of a pristine $\text{LiBO}_2$ pellet: (A) Overview cross-section under the protective tungsten (W) cap. (B, C) HRTEM and SAED images  show the pristine single crystal of a location (labeled with a green circle) in the subsurface region immediately underneath of the W cap.}
\label{figure_TEM_pristine}
\end{figure*}

\begin{figure*}
\centering
\includegraphics[width=\linewidth]{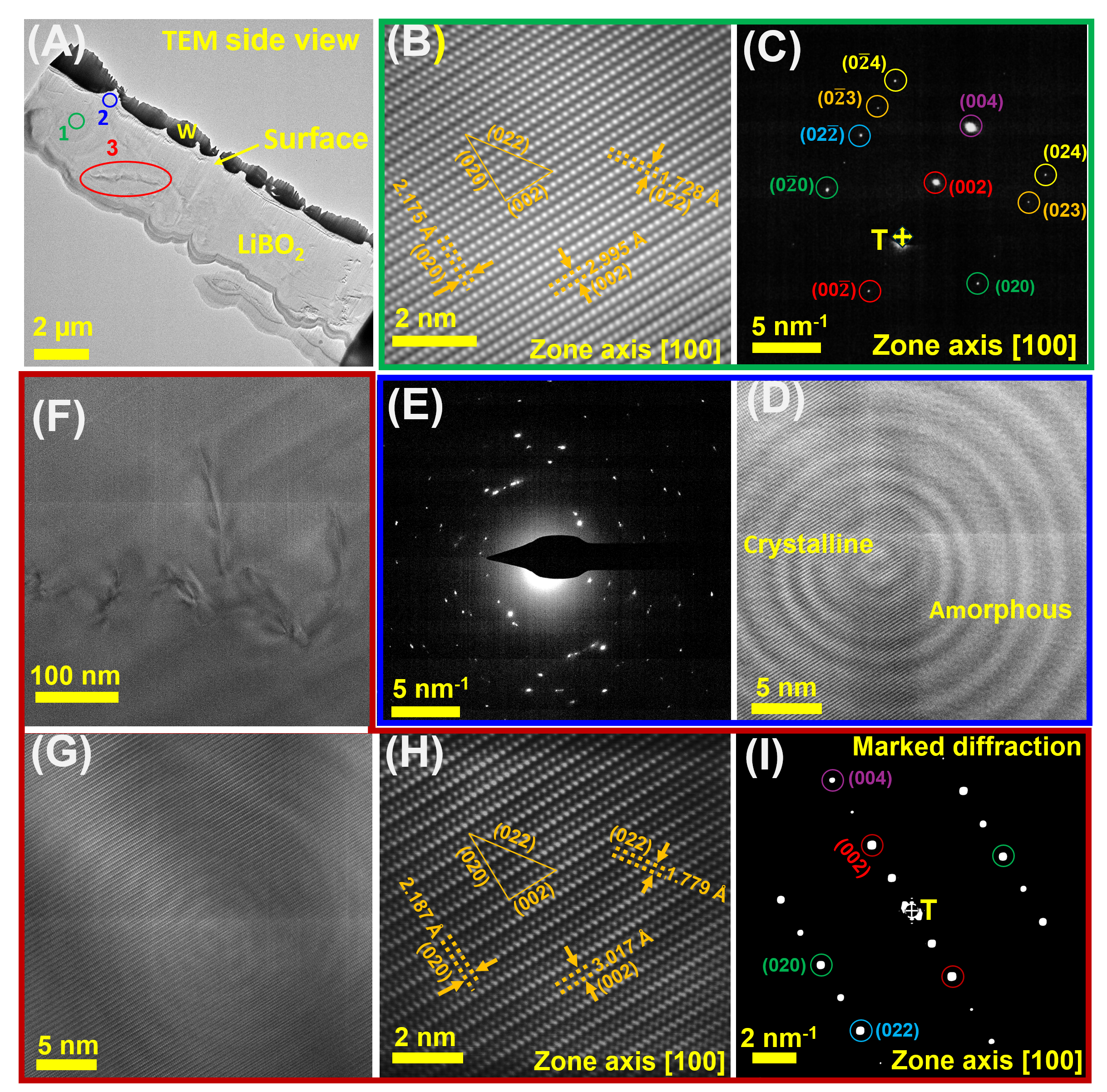}
\caption{Transmission electron microscopy analysis of a 10-min irradiated $\text{LiBO}_2$ crater cross-section: (A) Overview cross-section under the protective tungsten (W) cap revealing three distinct structural regions (Locations 1, 2, and 3). (B, C) HRTEM imaging and SAED at Location 1 showing the pristine single-crystal baseline. (D, E) HRTEM and SAED at Location 2 (crater mouth) showing a heterogeneous mixture of coexisting crystalline and amorphous phases. (F--I) HRTEM imaging and corresponding real/reciprocal space analyses at Location 3 (bulk interior) showing nuclear track penetration with single-crystal matrix preservation.}
\label{figure5_TEM}
\end{figure*}

To directly resolve the nanometer-scale structural and microstructural modification and have more insight into the nuclear-recoil-induced cratering, TEM analyses were conducted on cross-sectional lamellas extracted from a pristine and a 10-min irradiated $\text{LiBO}_2$ pellets, whose outcomes of our TEM analysis are shown in Figures \ref{figure_TEM_pristine} and \ref{figure5_TEM}, respectively. 

Because lithium compounds are highly prone to moisture-induced surface reconstruction in ambient air and electron-beam-induced radiolytic damage under standard high-voltage excitation ($\sim 200\text{ kV}$), sample preparation was performed via a customized cryo-Focused Ion Beam (cryo-FIB) protocol for cryo-TEM sample preparation and characterization, which is described in the Experimental Methods below. 

Figure \ref{figure_TEM_pristine} displays the TEM characterization of the unirradiated, pristine $\text{LiBO}_2$ reference sample. The cross-sectional overview image (panel \ref{figure_TEM_pristine}A) shows a uniform, dense cross-section prepared under a protective tungsten (W) mask. The original sample surface beneath the W layer is exceptionally flat and featureless, exhibiting no evidence of surface cratering, micro-voids, or localized topographic depressions. High-resolution TEM (HRTEM) imaging recorded from the interior of the subsurface grain at the sampled location (see the green circle on panel \ref{figure_TEM_pristine}B) demonstrates continuous, defect-free lattice fringes with uniform contrast and long-range periodicity, confirming the presence of a highly ordered crystalline matrix. Furthermore, the corresponding selected-area electron diffraction (SAED) pattern (panel \ref{figure_TEM_pristine}C) shows a clean, well-defined spot array characterized by sharp, discrete Bragg reflections devoid of diffuse halos, streaking, or spot-broadening. 

Figure \ref{figure5_TEM}A displays the low-magnification cross-sectional TEM overview of the milled crater region beneath the deposited W mask. Immediately below the protective W layer, the native sample surface exhibits an undulating profile characterized by sub-micrometer valleys and nanoscale depressions within the primary crater floor. Because the protective W cap conforms directly to the frozen top surface prior to ion milling, these nanoscale sub-craters provide direct physical proof that the micrometer-scale SEM crater is an emergent, coalesced structure resulting from multiple localized, overlapping nuclear interaction events.

To systematically map the structural and phase variations across the damage landscape, three distinct regions were examined: Location 1 (pristine bulk control, green), Location 2 (crater rim/mouth, blue), and Location 3 (bulk ion-track region, red).

At Location 1, positioned deep within an unaffected grain interior, HRTEM (panel \ref{figure5_TEM}B) and SAED (panel \ref{figure5_TEM}C) reveal an intact, highly ordered single-crystalline lattice. The absence of detectable lattice disorder or phase decomposition at Location 1 provides an important internal reference for assessing the structural modifications observed at Locations 2 and 3.

Location 2 (panels \ref{figure5_TEM}D,E), situated at the immediate surface and rim of the crater valley, exhibits a heterogeneous mixture of coexisting amorphous pockets and disordered crystalline domains. High-resolution TEM (HRTEM, panel \ref{figure5_TEM}D) directly shows nanoscale crystalline lattice fringes interspersed with featureless, highly disordered amorphous zones resulting from rapid surface thermal spike quenching. This structural heterogeneity is confirmed by the corresponding SAED pattern (panel \ref{figure5_TEM}E), which displays broad, diffuse halo rings characteristic of short-range amorphous disorder, superimposed on faint, broadened Bragg spots from residual disordered crystalline domains.

In contrast, Location 3 (panels \ref{figure5_TEM}F--I), located beneath the crater floor within the grain interior, captures the microstructural footprint of nuclear track penetration by energetic daughter products ($^{4}\text{He}$, $^{7}\text{Li}$, and $^{3}\text{H}$) generated via the afore-mentioned transmutation reactions [see Eqns (\ref{eq3}) and (\ref{eq4})]. The TEM overview (panel \ref{figure5_TEM}F) highlights linear defect contrast traversing the bulk matrix. High-resolution TEM imaging of this track region (panel \ref{figure5_TEM}G) reveals real-space concentric intensity rings, similar to those observed at Location 2 (panel \ref{figure5_TEM}D), whereas no comparable concentric contrast is evident in the pristine baseline at Location 1 (panel \ref{figure5_TEM}B). The spatial correlation of these features with the irradiated regions suggests that they may be associated with irradiation-induced structural perturbations surrounding the recoil-track region. However, the HRTEM contrast alone does not uniquely determine their origin. Possible mechanisms include localized lattice strain or bending, Moiré/interference effects from overlapping or slightly misoriented crystalline regions, and diffraction contrast associated with local lattice misorientation. Thus, the concentric contrast may provide evidence of localized structural modification at the corresponding locations (Locations 2 and 3), although its precise origin cannot be established from the present HRTEM images alone. Further investigation using dose-dependent imaging, complementary diffraction and STEM-based characterization, and comparison with simulated contrast would be valuable for elucidating the nature and origin of these features. Nevertheless, the persistence of crystalline lattice fringes in the surrounding material further indicates that any irradiation-induced disorder is spatially heterogeneous. Determining the relative contributions of strain, lattice misorientation, and imaging-related contrast will require complementary diffraction and/or local strain analysis.

To resolve the underlying matrix symmetry beneath this strain field, a targeted region of panel \ref{figure5_TEM}G was Fourier-filtered and denoised to yield the processed real-space HRTEM image (panel \ref{figure5_TEM}H) and its corresponding reciprocal-space fast Fourier transform (FFT) pattern (panel \ref{figure5_TEM}I) and its corresponding reciprocal-space fast Fourier transform (FFT) pattern (panel \ref{figure5_TEM}I). Crucially, the FFT in panel \ref{figure5_TEM}I displays a sharp, well-defined single-crystal spot array that directly mirrors the pristine SAED baseline at Location 1 (panel \ref{figure5_TEM}C). This confirms that despite the localized strain-induced contrast probably observed in panel \ref{figure5_TEM}G, the core bulk matrix hosting the recoil track retains a fully coherent single-crystal framework.

This stark contrast between the amorphized surface crater rim (Location 2, panels \ref{figure5_TEM}D,E) and the single-crystalline bulk track (Location 3, panels \ref{figure5_TEM}F--I) might be attributed to fundamental differences in heat dissipation and thermal relaxation dynamics between surface and bulk environments in an insulating oxide:

\textbf{Bulk Recrystallization via $4\pi$ Solid-Angle Heat Sinking (Location 3):} In an electrically insulating material such as $\text{LiBO}_2$, heat transport relies almost exclusively on phonon conduction ($k \approx k_{\text{ph}}$). Deep within the grain interior, a transient thermal spike generated along a daughter-particle recoil track is surrounded across a full $4\pi$ solid angle by the cold, ordered host matrix. This geometric configuration enables rapid radial dissipation of thermal energy via solid-state phonon scattering. Because the transiently molten track volume is enveloped on all sides by the undamaged lattice, the rapid thermal extraction is accompanied by immediate homoepitaxial re-solidification seeded by the surrounding single crystal, restoring long-range crystallographic order along the track trajectory and ensuring bulk XRD reflections remain sharp.
    
\textbf{Surface Quenching and Phonon Bottleneck (Location 2):} Conversely, at the free surface, solid-state heat sinking is geometrically restricted to a $2\pi$ solid angle. Thermal energy cannot conduct via phonons into the vacuum boundary, and radiative heat loss ($\propto \sigma T^4$) is orders of magnitude slower than solid-state phonon transport on picosecond timescales. This creates a localized heat-dissipation bottleneck that prolongs the thermal spike duration at the surface boundary, driving explosive surface atom ejection (sputtering) and localized melting. As this surface melt cools, the loss of stoichiometry through volatile species ejection, combined with the lack of an enclosing 3D crystalline template, prevents orderly epitaxial recovery, quenching the interface into a heterogeneous mixture of amorphous pockets and disordered crystalline domains.

Finally, it is worth correlating between PXRD/GIXRD and localized TEM/SAED outcomes. The localized amorphization observed at Location 2 (panels \ref{figure5_TEM}D,E) provides a vital complementary bridge to our macro-scale PXRD and GIXRD findings (Figure \ref{figure2_XRD}). While bulk XRD techniques register sharp, unbroadened Bragg peaks indicating that long-range monoclinic matrix symmetry is fully preserved, XRD is fundamentally an ensemble-averaging measurement over macroscopic sample volumes ($\sim 1\text{--}10\text{ mm}^3$). Because thermal-neutron-induced phase disruption is strictly confined to nanometer-scale surface crater rims and discrete recoil tracks, the total volume fraction of the amorphous/disordered phase remains well below the detection limit ($\sim 3\text{--}5\text{ vol}\%$) of standard X-ray diffraction. In contrast, localized SAED/FFT and HRTEM probe precise, nanometer-scale interaction volumes ($< 10^{-15}\text{ mm}^3$), enabling direct visualization of localized structural amorphization and defect tracks that remain otherwise invisible to bulk powder XRD.

In summary, the cryo-TEM analysis provides multi-scale experimental proof that: (1) thermal neutron absorption creates localized sub-micron craters that coalesce into the mesoscale craters observed in SEM; (2) surface regions undergo localized amorphization due to hindered heat dissipation during thermal spike quenching, which remains undetected in bulk XRD due to its extremely small volume fraction; and (3) bulk grain interiors undergo efficient epitaxial recrystallization, preserving long-range matrix single-crystallinity while hosting localized defect tracks. These nanometer-scale amorphous-crystalline boundaries and point-defect tracks directly account for the enhanced low-activation transport channels evaluated in subsequent conductivity analyses.

\subsection{Effects of Irradiation Dose on Structure and Microstructure of LiBO$_{2}$ \label{section3_2}}

\begin{table}[htbp]
\centering
\caption{Lattice parameters obtained from Rietveld refinement}
\label{table1_rietveld}
\small
\begin{tabular}{llcccccccc}
\toprule
& & \multicolumn{5}{c}{Monoclinic LiBO$_2$} & \multicolumn{3}{c}{Tetragonal Li$_2$B$_4$O$_7$} \\
\cmidrule(lr){3-7} \cmidrule(lr){8-10}
Method & Sample & $a$ (\AA) & $b$ (\AA) & $c$ (\AA) & $\beta$ ($^\circ$) & W\% & $a$ (\AA) & $c$ (\AA) & W\% \\
\midrule
PXRD & Raw & 5.83 & 4.34 & 6.44 & 114.96 & 97 & 9.43 & 10.28 & 3 \\
 & Pristine & 5.86 & 4.37 & 6.48 & 115.04 & 94 & 9.44 & 10.45 & 6 \\
 & 10 min & 5.83 & 4.35 & 6.45 & 114.84 & 98 & 9.47 & 10.30 & 2 \\
 & 60 min & 5.86 & 4.37 & 6.48 & 115.09 & 93 & 9.50 & 10.39 & 7 \\
 & 120 min & 5.84 & 4.35 & 6.46 & 115.05 & 92 & 9.48 & 10.28 & 8 \\
 & 12 h & 5.85 & 4.35 & 6.46 & 115.04 & 91 & 9.48 & 10.30 & 9 \\
GIXRD & Pristine & 5.85 & 4.35 & 6.46 & 115.05 & 94 & 9.48 & 10.30 & 6 \\
 & 10 min & 5.84 & 4.35 & 6.46 & 115.06 & 96 & 9.48 & 10.28 & 4 \\
\midrule
\multicolumn{2}{l}{Average} & 5.84 & 4.35 & 6.46 & 115.02 & 94 & 9.47 & 10.32 & 6 \\
\multicolumn{2}{l}{Standard Deviation} & 0.01 & 0.02 & 0.02 & 0.18 & 3 & 0.04 & 0.13 & 3 \\
\bottomrule
\end{tabular}
\end{table}

\begin{figure}[htbp]
    \centering
    \includegraphics[width=\linewidth]{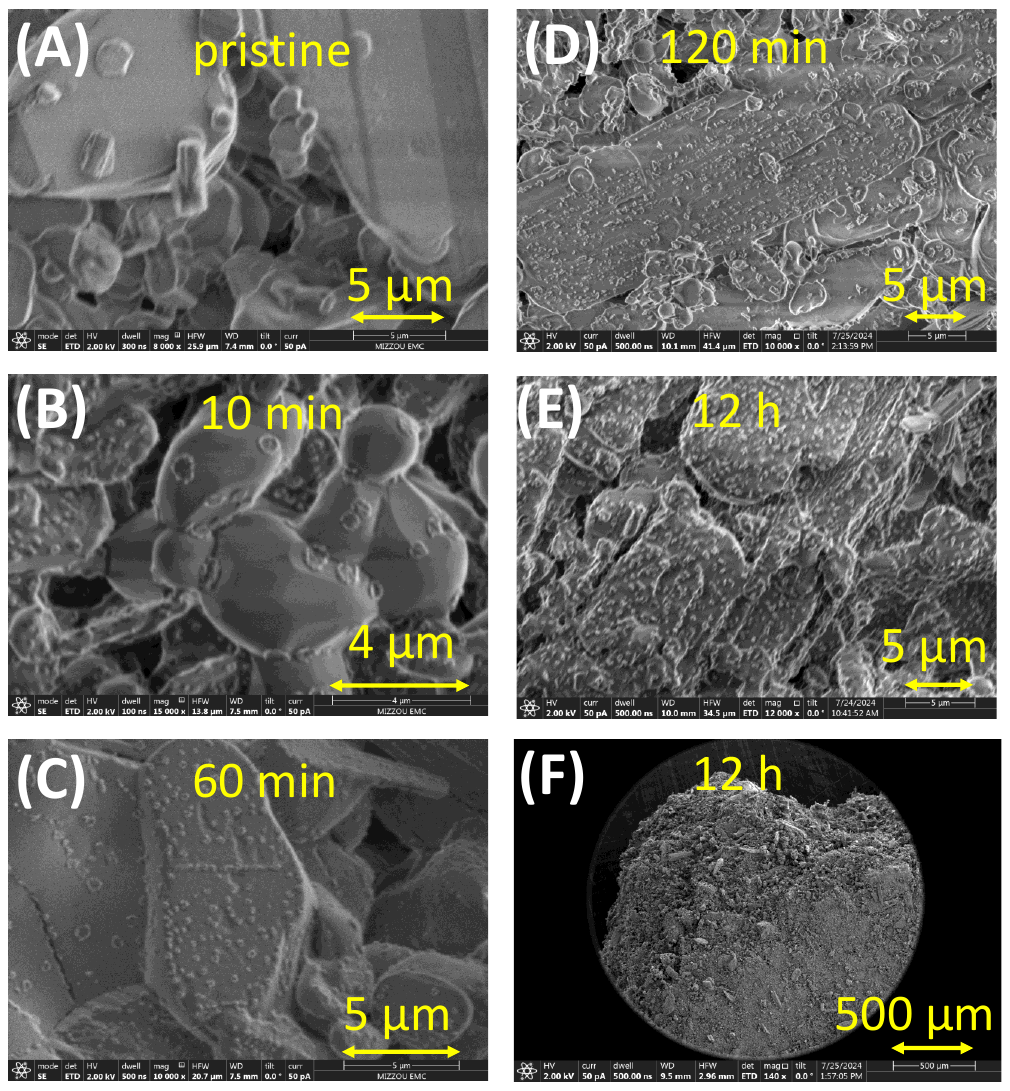}
    \caption{Dose-dependent Microstructure of LiBO$_{2}$.}
    \label{figure5_SEM_all}
\end{figure}

The evolution of the LiBO$_{2}$ system under increasing thermal neutron exposure reveals a clear correlation between irradiation dose and microstructural modification, even as the long-range crystallographic order remains remarkably resilient. 

As one can see in Fig.~\ref{figure2_XRD}, the characteristic PXRD peaks for monoclinic LiBO$_{2}$ persist across all dosage levels—from the pristine state through 10 minutes, 60 minutes, 120 minutes, and up to 12 hours of irradiation—indicating that the bulk crystal framework does not undergo notable modification of the crystallographic order (see the inset of Figure \ref{figure2_XRD}). As a result, the outcome of the Rietveld refinement for all of the diffraction patterns (see Table \ref{table1_rietveld}) shows that the lattice parameters and weight fractions of the two crystal phases slightly fluctuate around their average values [major monoclinic LiBO$_{2}$ ($94 \pm 3$ \%): $a = 5.84 \pm 0.01\text{ \AA}$, $b = 4.35 \pm 0.02 \text{ \AA}$ and $c = 6.46 \pm 0.02 \text{ \AA}$, $\alpha=\gamma = 90^{o}$ and $\beta= 115.02 \pm 0.18^{\text{o}}$; and minor tetragonal Li$_{2}$B$_{4}$O$_{7}$ ($6 \pm 3$ \%): $a = b = 9.47 \pm 0.04\text{ \AA}$ and $c = 10.32 \pm 0.13 \text{ \AA}$].  In other words, no clear dose-dependent trend is found from the lattice parameters obtained from the refinement, suggesting that the monoclinic lattice remains largely intact, even under prolonged exposure to thermal neutrons. 

However, a detailed examination of the microstructure in Figure \ref{figure5_SEM_all} (panels A–E) highlights a progressive accumulation of surface and interface modifications, which is summarized as follows.
    
\textbf{Pristine to Short-Term Exposure (0–10 min)}: As shown in Figure \ref{figure3_SEM_pristine} and panel A of Figure \ref{figure5_SEM_all}, the pristine sample starts as a dense polycrystalline network with well-defined grains and residual porosity. After 10 minutes (Figure \ref{figure4_SEM_10min_irradiated} and panel B of Figure \ref{figure5_SEM_all}), the initial effects of neutron capture by $^{10}$B and $^{6}$Li manifest as localized crater-like features and enlarged pore mouths. These craters are non-uniformly distributed, preferentially forming at triple junctions and grain boundaries where the lattice is naturally more susceptible to displacement cascades. A more detailed discussion of the microstructural contrast between pristine and 10-min irradiated pellets can be seen in the previous section (Section \ref{section3_1}).

\textbf{Intermediate Exposure (60–120 min)}: With increased irradiation time to 60 minutes (panel C of Figure \ref{figure5_SEM_all}) and 120 minutes (panel D of Figure \ref{figure5_SEM_all}), there is a noticeable increase in the density and size of these surface craters. Localized energy deposition from energetic particles (e.g., alpha particles, tritons, lithium ions, etc.) leads to more extensive material ejection and the interconnection of subsurface pores. This suggests that while the grains themselves remain structurally coherent, the ``plumbing'' of the material—its grain boundaries and outgassing pathways—is being significantly modified.

\textbf{Extended Exposure (12 h)}: At the highest irradiation dose (12 h; panels E and F of Figure~\ref{figure5_SEM_all}), the cumulative effects of neutron-induced defects, helium production, and thermal spikes lead to pronounced mechanical degradation. Although Figure~\ref{figure2_XRD} confirms that the LiBO$_{2}$ phase is largely preserved, the pellet exhibits macroscopic fracture (panel F), indicating loss of structural integrity. This transition from localized surface cratering to bulk failure suggests that irradiation-induced stresses eventually approach or exceed the cohesive strength of the polycrystalline framework. An order-of-magnitude estimate of helium bubble pressurization supports this interpretation: for a nanoscale bubble of radius $r \sim 1$--5~nm and atomic density $n_0 \sim 3 \times 10^{28}~\text{m}^{-3}$, a helium filling fraction $f \sim 0.1-1$ gives $n_{\text{He}} \sim f n_0 \sim 10^{27}$--$10^{28}~\text{m}^{-3}$, and, under transient thermal spike conditions ($T \sim 10^{4}$~K), the pressure can be estimated as $p \approx n_{\text{He}} k_{\text{B}} T \sim 10^{8}$--$10^{9}$~Pa ($0.1-1$~GPa). This estimate is conservative, as confinement and non-ideal effects can further elevate pressures into the multi-GPa regime. Such stress levels are comparable to the tensile and grain-boundary strengths of brittle oxides, implying that helium bubble growth, coalescence, and associated stress concentration—particularly at microstructural heterogeneities—can drive crack initiation and propagation, ultimately resulting in macroscopic fracture under prolonged irradiation.

In summary, the irradiation dose acts as a precision tool for defect engineering at lower exposures (up to 120 min) by selectively modifying ion transport channels at grain boundaries. However, excessive doses lead to a ``saturation" of defects that results in the mechanical breakdown of the material, marking the upper dose limit that can be optimally set for effective neutron-induced engineering of ion transport in LiBO$_{2}$ without compromising the the integrity of  whole solid pellets.


\subsection{Effects of Irradiation Dose on Ionic Conductivity \label{section3_3}}

As described in the Introduction, the primary objective of this work is to demonstrate that thermal neutron irradiation can serve as a controllable defect-engineering strategy to selectively modify ion transport in LiBO$_{2}$, leading to its enhanced ionic conductivity. Specifically, neutron-induced defects (e.g., Li and B vacancies) are expected to alter ionic migration pathways and thereby enhance the overall ionic conductivity (see also Fig.~\ref{figure1_working_principle}). This section presents and discusses the effects of irradiation time on ion transport in grains and grain boundaries of polycrystalline LiBO$_{2}$ pellets subjected to thermal neutron radiation.

\begin{figure*}
    \centering
    \includegraphics[width= \linewidth]{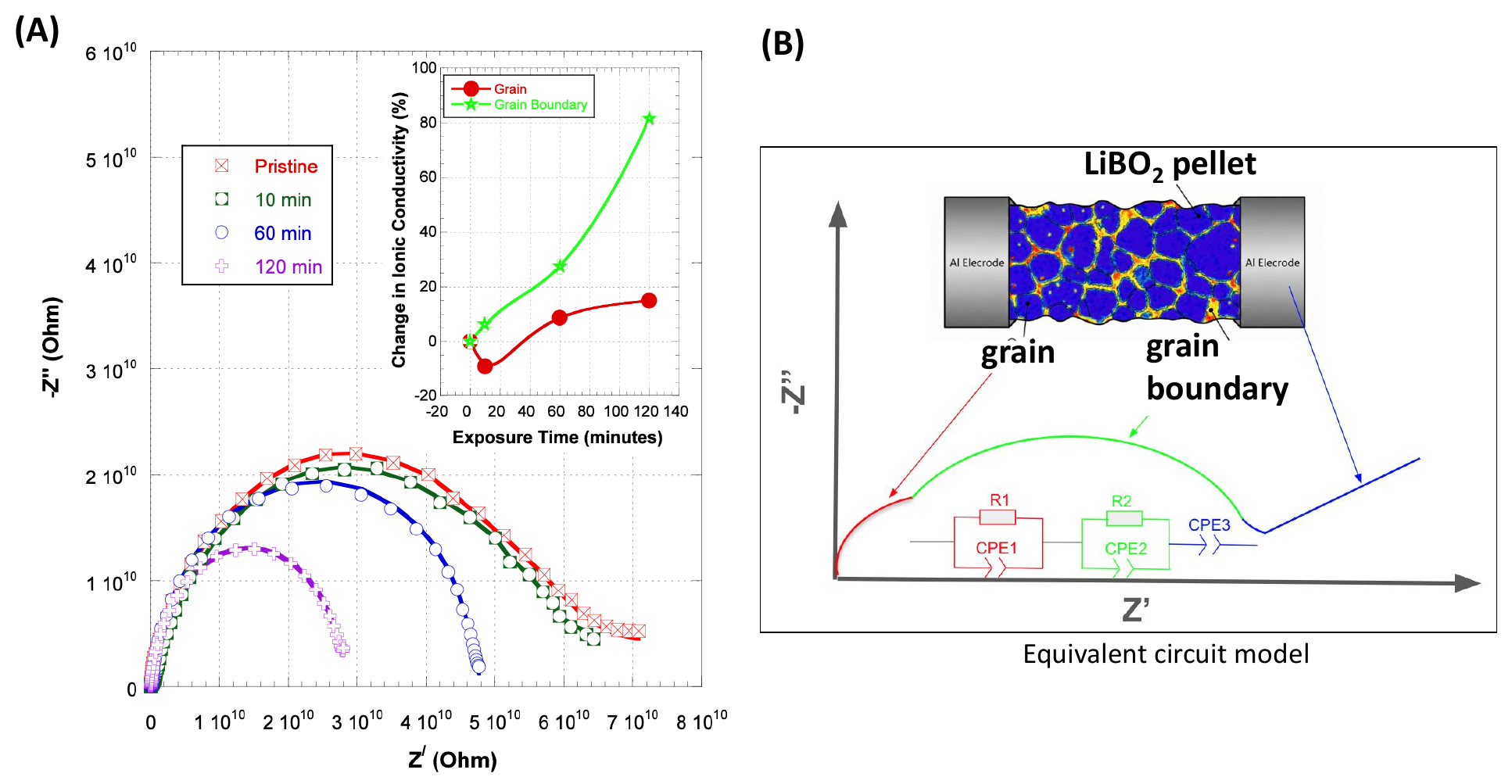}
    \caption{Dose-dependent EIS and ionic conductivity of LiBO$_{2}$.}
    \label{figure6_EIS_data}
\end{figure*}

To quantify these effects, electrochemical impedance spectroscopy (EIS) measurements were performed at room temperature on pristine and irradiated pellets using symmetric blocking aluminum electrodes (upper inset, right panel of Fig.~\ref{figure6_EIS_data}). The measurements were conducted over the frequency range from 1~mHz to 1~MHz, and the resulting Nyquist plots are shown in Fig.~\ref{figure6_EIS_data}A. The experimental data (symbols) were fitted (solid lines) using an appropriate equivalent circuit model (lower inset, Fig.~\ref{figure6_EIS_data}B), enabling extraction of the ionic conductivities associated with the grain interiors ($\sigma_{1}$) and the grain boundaries ($\sigma_{2}$).

 The impedance response of a polycrystalline LiBO$_{2}$ pellet \cite{Hirose2019}—consistent with typical SSIC behavior \cite{Thomas_Defferriere_2022,Thomas_Defferriere_2024,Thomas_Defferriere_2025,Hui2007,JianFangWua_and_Xin_Guo_2016,Ban_Choi_2001,Tomasz_Polczyk_2021,Yanlin_Zhu_2020}—can be interpreted using an equivalent electrical circuit consisting of two resistor–constant phase element (R--CPE) units connected in series, followed by an additional CPE in series \cite{Thomas_Defferriere_2022,Thomas_Defferriere_2024,Thomas_Defferriere_2025,Hirose2019,Hui2007,JianFangWua_and_Xin_Guo_2016,Ban_Choi_2001,Tomasz_Polczyk_2021,Yanlin_Zhu_2020,Lazanas2023,PoojaVadhva2021}:

\begin{equation}
(R_1 \parallel \text{CPE}_1) 
- (R_2 \parallel \text{CPE}_2) 
- \text{CPE}_3.
\end{equation}

Here, the subscript ``1'' corresponds to the high-frequency response of the grain (bulk) interior, while ``2'' represents the medium-frequency contribution of grain boundaries. The third element, CPE$_3$, accounts for low-frequency electrode polarization effects (see schematic Nyquist plot in Fig.~\ref{figure6_EIS_data}B).

The grain ionic conductivity is determined from the extracted grain resistance $R_1$ as

\begin{equation}
\sigma_1 = \frac{L}{R_1 A},
\label{eqsigma_1}
\end{equation}

\noindent where $L = 0.3$~cm and $A \approx 2$~cm$^{2}$ are the thickness of the pellet and its cross-sectional area, respectively. 

It is noted that the constant phase element (CPE) was introduced to model non-ideal capacitive behavior commonly observed in polycrystalline materials. Its complex impedance is defined as

\begin{equation}
Z_{\text{CPE}} 
= \frac{1}{Q (j\omega)^n} 
= \frac{1}{Q \omega^n} 
\exp\left(-j \frac{n\pi}{2}\right),
\end{equation}

\noindent hence, the complex impedance of a $R$-CPE in parallel is given by

\begin{equation}
    Z_{R\parallel \text{CPE}}(\omega) = \frac{1}{\dfrac{1}{R} + Q (j\omega)^n},
\end{equation}

\noindent which can be separated into its real and imaginary components as
\begin{align}
    Z'(\omega) &= \frac{R \big( 1 + (RQ\,\omega^n)\cos(\tfrac{n\pi}{2}) \big)}
    {1 + 2(RQ\,\omega^n)\cos(\tfrac{n\pi}{2}) + (RQ\,\omega^n)^2}, \\
    Z''(\omega) &= -\frac{R (RQ\,\omega^n)\sin(\tfrac{n\pi}{2})}
    {1 + 2(RQ\,\omega^n)\cos(\tfrac{n\pi}{2}) + (RQ\,\omega^n)^2},
\end{align}

\noindent where $j$ is the imaginary unit, $\omega$ is the angular frequency, $Q$ is the CPE constant, and $0 < n \le 1$ is a dimensionless exponent describing the deviation from ideal capacitive behavior (with $n = 1$ corresponding to an ideal capacitor).

The effective capacitance associated with a CPE--resistor pair can be estimated as

\begin{equation}
C = \left(R^{1-n} Q \right)^{1/n},
\end{equation}

\noindent from which the grain-boundary ionic conductivity can be determined as

\begin{equation}
\sigma_2 = \frac{L}{R_2 A}\frac{C_1}{C_2}=\frac{L}{R_2 A}\frac{\left(R_1^{1-n_1} Q_1 \right)^{1/n_1}}{\left(R_2^{1-n_2} Q_2 \right)^{1/n_2}}.
\end{equation}

In addition, for irradiated samples, the corresponding conductivities are denoted as $\sigma_{1,\text{irr}}$ and $\sigma_{2,\text{irr}}$, while those of pristine samples are $\sigma_{1,\text{pri}}$ and $\sigma_{2,\text{pri}}$. The relative conductivity change was calculated as

\begin{equation}
\Delta \sigma_i (\%) 
= \frac{\sigma_{i,\text{irr}} - \sigma_{i,\text{pri}}}
{\sigma_{i,\text{pri}}} \times 100, 
\qquad i = 1,2.
\label{percentchangeconductivity}
\end{equation}

\begin{figure}
    \centering
    \includegraphics[width= \linewidth]{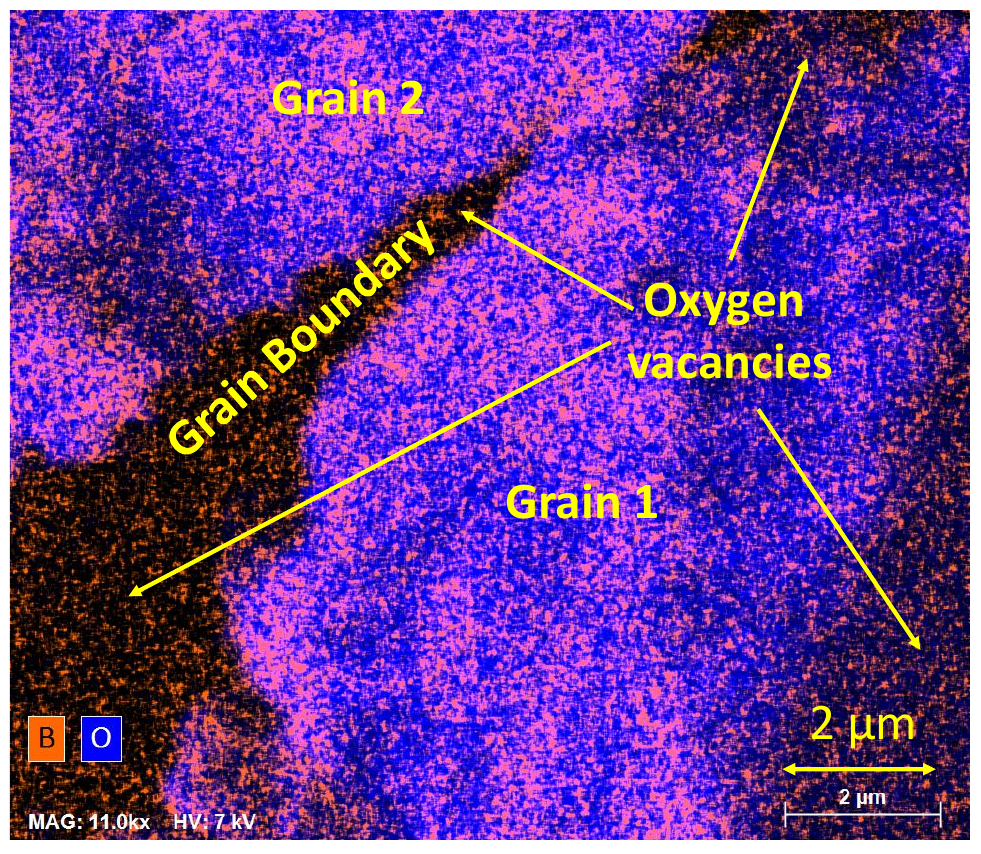}
    \caption{Energy-Dispersive X-ray Spectroscopy (EDS) of the surface of a pristine LiBO$_2$ pellet.}
    \label{EDS}
\end{figure}

All fitting parameters obtained from the EIS analysis of the pristine and irradiated pellets are provided in Table \ref{table2_eis}. For the pristine pellet, the fitted grain resistance is $R_{1} = 1.0 \times 10^{7}\text{ }\Omega$ (10 M$\Omega$), which is approximately three orders of magnitude smaller than the grain-boundary resistance, $R_{2} = 5.1 \times 10^{10}\text{ }\Omega$ (51 G$\Omega$). This large difference is attributed to the formation of a space-charge region at the grain boundaries arising from the accumulation of positively charged oxygen vacancies, as suggested by our Energy-Dispersive X-ray (EDS) spectroscopy data (see Figure \ref{EDS}). Such grain-boundary blocking effects have been observed in some of SSICs. For example, Wu and Guo \cite{JianFangWua_and_Xin_Guo_2016} reported an even stronger grain-boundary blocking effect (approximately four orders of magnitude) in the perovskite Li$_{3x}$La$_{2/3-x}$$\square$$_{1/3-2x}$TiO$_{3-\delta}$ (here, $3x = 0.263$ and $\square$ denotes a vacancy site), where the large disparity between grain and grain-boundary ionic conductivities was attributed to positively charged space-charge layers at grain boundaries, which can be partially neutralized by dynamically introduced electrons. Experimental strategies to mitigate grain-boundary blocking in SSICs include increasing the annealing temperature during solid-state synthesis \cite{Ban_Choi_2001}, deliberate grain-boundary engineering \cite{Tomasz_Polczyk_2021}, grain-boundary amorphization \cite{Yanlin_Zhu_2020}, and irradiation with ionizing radiation [e.g., ultraviolet (UV) or gamma rays] \cite{Thomas_Defferriere_2022,Thomas_Defferriere_2024,Thomas_Defferriere_2025}.

\begin{table}[htbp]
\centering
\caption{Fitting parameters of EIS data modeling for pristine and irradiated LiBO$_2$.}
\label{table2_eis}
\small
\begin{tabular}{lcccccccc}
\toprule
& \multicolumn{3}{c}{Grain} & \multicolumn{3}{c}{Grain Boundary} & \multicolumn{2}{c}{Electrode} \\
\cmidrule(lr){2-4} \cmidrule(lr){5-7} \cmidrule(lr){8-9}
Sample & $Q_1$ ($10^{-10}$) & $n_1$ & $R_1$ ($10^6\,\Omega$) & $Q_2$ ($10^{-10}$) & $n_2$ & $R_2$ ($10^9\,\Omega$) & $Q_3$ ($10^{-8}$) & $n_3$ \\
\midrule
pristine & 0.32 & 0.98 & 10  & 0.80 & 0.90 & 51.0 & 0.60 & 0.7  \\
10 min   & 0.55 & 0.98 & 11  & 0.75 & 0.90 & 48.0 & 0.60 & 0.7  \\
60 min   & 0.25 & 0.98 & 9.2 & 0.97 & 0.98 & 40.0 & 0.60 & 1.2  \\
120 min  & 0.82 & 0.98 & 8.7 & 0.95 & 0.95 & 28.1 & 3.1  & 0.60 \\
\bottomrule
\end{tabular}
\end{table}

In the present work, we demonstrate for the first time that thermal neutron transmutation can serve as an alternative strategy to mitigate grain-boundary blocking. During thermal neutron irradiation, the transmutation reactions of $^{6}$Li and $^{10}$B generate cascades and subcascades of atomic displacements, accompanied by the emission of energetic particles and gamma photons. These processes promote the redistribution of atoms and electrons, which likely leads to partial neutralization of the space charge in the grain-boundary region, thereby reducing the impedance to Li$^{+}$ ion transport across grain boundaries. In particular, electrons displaced or liberated by gamma photons produced as by-products of the $^{10}$B transmutation [See the second nuclear reaction in Eqn. (\ref{eq3})] may contribute to this neutralization mechanism (see Fig.~\ref{figure1_working_principle}). The reversible neutralization of grain-boundary space charge with electrons excited by gamma and UV photons has previously been reported by Defferriere et al. \cite{Thomas_Defferriere_2022,Thomas_Defferriere_2024}. In contrast, in the present study we observe that the structural, microstructural, and electrical modifications of LiBO$_{2}$ pellets occur in a nonreversible and dose-dependent manner. Specifically, compared with the pristine pellets ($R_{2} = 51.0\text{ G}\Omega$), pellets irradiated with thermal neutrons for 10, 60, and 120 minutes exhibit progressively reduced grain-boundary resistance values of 48.0, 40.0, and 28.1 $\text{G}\Omega$, corresponding to reductions of 5.9\%, 21.6\%, and 44.9\%, respectively. It is noted that the pellets irradiated for 12 h fractured due to helium buildup resulting from transmutation reactions and therefore could not be characterized by EIS. 

In contrast to the monotonic reduction observed for the grain-boundary resistance, the effect of irradiation dose on the grain (bulk) resistance exhibits a gradual and nonmonotonic behavior. Relative to the pristine pellet ($R_{1}=10\text{ M}\Omega$), the grain resistance values for samples irradiated with thermal neutrons for 10, 60, and 120 minutes are $11$, $9.2$, and $8.7\text{ M}\Omega$, corresponding to relative changes of $+10\%$, $-8\%$, and $-13\%$, respectively (where the sign indicates an increase or decrease with respect to the pristine value).

The initial increase in $R_{1}$ after 10 minutes of irradiation suggests that, at low defect concentrations, irradiation-induced lattice defects such as Li and B vacancies remain spatially isolated and therefore act predominantly as scattering or trapping centers for mobile Li$^{+}$ ions, slightly impeding intra-grain ion transport. As the irradiation dose increases, however, the defect population becomes sufficiently dense for these vacancy sites to form a connected network that enables vacancy-mediated Li$^{+}$ diffusion across the grain interior. This transition from isolated defects to an interconnected defect landscape can facilitate long-range percolative Li$^{+}$ transport, thereby reducing the bulk resistance.

This interpretation is consistent with the conclusions drawn by Deck and Hu \cite{Deck_Hu_2023}, who showed that entropy-driven disorder enhances Li$^{+}$ mobility primarily by weakening local Li$^{+}$–anion interactions; however, a substantial increase in macroscopic ionic conductivity emerges only when such local disorder becomes integrated into thermodynamically stable, extended structures that support percolating diffusion pathways. The present experimental observation is also consistent with density functional theory predictions \cite{HaMNguyen2025a,HaMNguyen2025b}, which indicate that high concentrations of boron vacancies in LiBO$_2$ can significantly reduce the migration energy barrier for Li$^{+}$ ions via a vacancy-mediated diffusion mechanism. Importantly, these lattice vacancies are energetically unfavorable under equilibrium synthesis conditions and therefore difficult to achieve through conventional defect-engineering approaches such as solid-state reactions. Thermal neutron irradiation thus provides a unique non-equilibrium strategy for generating such defects, enabling controlled modification of the ion transport landscape within the crystal lattice.

Finally, Fig. \ref{figure6_EIS_data}A shows the percentage changes in the ionic conductivities obtained from Eqn. (\ref{percentchangeconductivity}) versus the irradiation time, illustrating the effects of the irradiation dose on the ionic conductivity. Clearly, the ionic conductivity is increased by nearly 20\% for the grains and more than 80\% for the grain boundaries.

\section{Conclusions \label{conclusions}}

In summary, we provide the first experimental demonstration that high-flux thermal neutrons delivered at Beam Port E of the University of Missouri Research Reactor (MURR) can be utilized to selectively induce nuclear transmutation reactions in the strong neutron-absorbing isotopes $^{10}$B and $^{6}$Li. These isotopes are intrinsically and homogeneously incorporated into the LiBO$_2$ lattice at their natural abundances ($\sim$19.9\% for $^{10}$B and $\sim$7.5\% for $^{6}$Li). From a processing perspective, irradiation at the beam port enables significantly higher sample throughput and operational flexibility, offering a scalable pathway for materials processing without requiring reactor shutdown for sample retrieval~\cite{JustynaMinkiewicz2023}.

The neutron-capture reactions in Eqns. (\ref{eq3}) and (\ref{eq4}) produce energetic charged particles that generate lattice defects, primarily in the form of vacancies, while largely preserving the long-range crystallographic framework of the host material. Concurrently, gamma photons emitted during the $^{10}$B transmutation process can induce secondary electronic excitations, leading to partial mitigation of irradiation-induced damage via electronic stopping and enhanced recombination processes. These radiation-induced electronic effects may also contribute to the partial screening of space-charge regions associated with positively charged oxygen vacancies, particularly at grain boundaries. As a result, the ion transport landscape is modified, yielding measurable enhancements in ionic conductivity of approximately $20\%$ within grains and more than $80\%$ across grain boundaries.

Beyond the specific case of LiBO$_2$, this work establishes a generalizable framework for neutron-driven defect engineering in polycrystalline solid-state ionic conductors (SSICs). In particular, thermal neutron irradiation enables access to non-equilibrium defect configurations that are not readily attainable through conventional thermochemical synthesis methods. This approach provides a strategy to simultaneously modulate both enthalpic and entropic contributions to ion transport, thereby complementing existing chemical doping and microstructural engineering strategies. The methodology introduced here holds promise for advancing the design of high-performance solid-state ionic materials for energy storage and conversion applications. Further systematic investigations across LiBO$_2$ and other SSIC systems are warranted to fully elucidate the interplay between neutron-induced defect populations, microstructure, and transport properties, and to establish irradiation as a controllable and scalable materials engineering tool.

\section{Experimental Methods and Materials \label{methods}}

Polycrystalline LiBO$_{2}$  pellets (coin-shaped, 1.6 cm in diameter and 0.3 cm in thickness) were synthesized from equal amounts of 0.5 grams of commercial anhydrous monoclinic LiBO$_{2}$ powder (Thermo, CAS No. 13453-69-5, Cat. No. 12591) by hot pressing (at 200$^{\circ}$C under a load of 2.5 tons) followed by sintering in air for 48 hours at 663 $^{\text{o}}$C (75\% of the melting point of  monoclinic LiBO$_{2}$ phase). This heat treatment was carried out to fuse LiBO$_{2}$ fine grains together into dense polycrystalline solid pellets without full melting them while relieving internal stresses and improving the crystallinity of sintered gains. Afterward, the post-sintered pellets were stored in an inert gas glove box filled with either argon or helium gas.

Thermal neutron irradiation was carried out at Beam Port E of MURR using a collimated thermal neutron beam (5-100 meV) with a flux of $\sim10^{9}$ neutrons$\cdot$cm$^{-2}\cdot$s$^{-1}$ \cite{JohnBrockmana1,JohnBrockmana2,JohnBrockmana3}. Irradiations were performed outside the reactor core under controlled dose conditions. Pristine and irradiated pellets were characterized using PXRD on the aforementioned Rigaku Ultima IV X-ray diffractometer, grazing incidence X-ray diffraction (GIXRD) on a Philips X-Pert diffractometer, scanning electron microscopy (SEM) on a Thermo Scientific$^{\text{TM}}$ Scios$^{\text{TM}}$ 2 DualBeam$^{\text{TM}}$ Sy Focused Ion Beam Scanning Electron microscope, focused ion beam scanning electron microscopy (FIB-SEM) on a Thermo Scientific$^{\text{TM}}$ Helios$^{\text{TM}}$ 5 Hydra UX DualBeam microscope,  and electrochemical impedance spectroscopy (EIS) measured using a Reference 3000 and Gamry's EIS300 software  and analyzed using Gamry’s Echem Analyst software. Comparative analysis of these data was used to evaluate irradiation-induced changes of the crystal structure, microstructure, and ion transport of LiBO$_{2}$. 

Cross-sectional lamellae were prepared using the Thermo Scientific$^{\text{TM}}$ Helios$^{\text{TM}}$ 5 Hydra UX DualBeam's plasma focused ion beam and Thermo Fisher Scientific$^{\text{TM}}$ Aquilos$^{\text{TM}}$ 2 DualBeam microscopes. Initial steps at Hydra used Ar and Xe gasses with a 30kV accelerating voltage for deposition of protective cap and milling of the trenches/J-cut. Then, the sample was lifted out and transferred to a copper half grid. Final steps of lamella polishing were performed at Aquilos 2 at liquid nitrogen temperatures with 30 kV, 8 kV and 5 kV using Ga ion beam. Aquilos microscope has a quick loader and a transfer station to minimize the ice formation during the transfer and keeping the sample inside the liquid nitrogen for most of the time to prevent oxidation. 
Transmission electron microscopy imaging and analysis was performed at Thermo Fisher Scientific Spectra 300 STEM at 300 kV accelerating voltage and with ~ 2-3 e-/px/s current. After the cryo station in Aquilos, the half-grid with lamella was transferred to precooled Gatan 915 double tilt cryo holder and inserted inside TEM. Typical temperature of operation of Gatan 915 ranges between 
-170$^{\text{o}}$C and -184$^{\text{o}}$C. After insertion, the holder with sample was left in microscope over 1–2 hours to stabilize and minimize the drift during imaging. Electron diffraction was done with <1pA current at 580 mm camera length with 20- and 40-micron apertures, selecting the area of interest. Energy dispersive X-ray spectroscopy elemental mapping was done at large field of view with 190 pA current to maximize the signal to noise ratio and to minimize beam damage done to the sample. High resolution imaging was done using Gatan K3 camera with a resolution of 3456$\times$3456 pixels. Acquisition was done in series with drift correction and averaging of 100 frames with 4 second total exposure.

\section*{Data Availability}

It is officially stated that the raw and reduced data that support the findings of this study are available from the corresponding authors upon reasonable request.

\section*{Author Contributions}
H.M.N and C.D.Z are the authors who contributed equally to the work. Principal Investigators (C.W, J.G., Y.X, T.W.H.); Conceptualization (C.W, J.G., Y.X, T.W.H., H.M.N); Sample preparation, X-ray diffraction and Rietveld (H.M.N, C.D.Z., N.N., T.W.H); SEM and FIB-SEM (C.D.Z., D.S, H.M.N); TEM (H M.N., M.S., T.P., T.H.T); EIS (H.M.N., C.D.Z., Z.R.V, B.S.), Thermal-neutron Irradiation (C.D.Z., J.G., T.W.H., H.M.N.); Data presentation and manuscript writing (H.M.N.); manuscript reviewing and editing (all authors). The authors declare no conflict of interest. All authors discussed, reviewed, edited, agreed, and were informed of the publication of the manuscript.

\section{Acknowledgements}

This work was funded by the University of Missouri Materials Science and Engineering Institute (MUMSEI) Grant No. CD002339, and the University of Missouri Research Reactor (MURR).

\end{document}